\RequirePackage{fix-cm}

\documentclass[smallextended]{svjour3}       %
\smartqed 
\usepackage{cite}
\usepackage{graphicx}
\usepackage{color}
\usepackage{caption}
\usepackage{subfig}
\usepackage[colorlinks,linkcolor=red,anchorcolor=blue,citecolor=green]{hyperref}
\usepackage{booktabs}
\usepackage{threeparttable}
\usepackage{float} 
\usepackage{multirow}

\usepackage{rotating}
\usepackage{array}
\usepackage{graphicx}
\usepackage{longtable,booktabs}
\usepackage{lscape}

\usepackage[T1]{fontenc}
\usepackage[utf8]{inputenc}
\usepackage{authblk}
\usepackage[misc]{ifsym}

\usepackage{amsmath}

\date{} 

\begin{document}

\title{A State-of-the-art Survey of U-Net in Microscopic Image Analysis: from Simple Usage to Structure Mortification}


\author{{Jian Wu}    \textsuperscript{a}   \and
{Wanli Liu}    \textsuperscript{a}  \and
        {Chen Li}   \textsuperscript{a,~\Letter} \and
        {Tao Jiang}    \textsuperscript{b}   \and
        {Islam Mohammad Shariful}    \textsuperscript{c}   \and
        {Hongzan Sun}    \textsuperscript{d}   \and
        {Xiaoqi Li}    \textsuperscript{a}   \and
        {Xintong Li}    \textsuperscript{a}   \and
        {Xinyu Huang}    \textsuperscript{e}   \and
        {Marcin Grzegorzek}    \textsuperscript{e}  
}

\institute{
    \begin{itemize}
      \item[\textsuperscript{\Letter}] {Corresponding author} \\
            \email{lichen201096@hotmail.com}
      \at
      \item[\textsuperscript{a}] Microscopic Image and Medical Image Analysis Group, 
      College of Medicine and Biological Information Engineering, 
      Northeastern University, China
      \item[\textsuperscript{b}] School of Control Engineering, 
      Chengdu University of Information Technology, China
      \item[\textsuperscript{c}] College of Software Engineering, Northeastern University, China
      \item[\textsuperscript{d}] Shengjing Hospital of China Medical University, China
\item[\textsuperscript{e}]Institute for Medical Informatics, University of Luebeck, Germany      
     \item[\textsuperscript{}]
    \end{itemize}
}

\authorrunning{J. Wu et al.: U-Net in Microscopic Image Analysis}

\titlerunning{J. Wu et al.: U-Net in Microscopic Image Analysis}        

\maketitle
\begin{abstract}
Image analysis technology is used to solve the inadvertences of artificial traditional 
methods in disease, wastewater treatment, environmental change monitoring analysis and 
convolutional neural networks (CNN) play an important role in microscopic image analysis. 
An important step in detection, tracking, monitoring, feature extraction, modeling and 
analysis is image segmentation, in which U-Net has increasingly applied in microscopic 
image segmentation. This paper comprehensively reviews the development history of U-Net, 
and analyzes various research results of various segmentation methods since the emergence 
of U-Net and conducts a comprehensive review of related papers. First, This paper has 
summarizes the improved methods of U-Net and then listed the existing significances of 
image segmentation techniques and their improvements that has introduced over the years. 
Finally, focusing on the different improvement strategies of U-Net in different papers, 
the related work of each application target is reviewed according to detailed technical 
categories to facilitate future research. Researchers can clearly see the dynamics of 
transmission of technological development and keep up with future trends in this 
interdisciplinary field.

\end{abstract}

\section{Introduction}
\label{intro}
\subsection{Background Knowledge of Microscopic Images }
Microscopic image refers to the image use to see in a microscope~\cite{Wu-2010-MIP}. 
There are many types of microscopes: The general magnification of the optical microscope 
is generally 1500-2000 times~\cite{Zenhausern-1994-ANO,Toledo-1992-NDS,Inouye-1994-NSO}, 
which has never exceeded 2000 times. However, the maximum magnification of the electron 
microscope exceeds three million times, such as transmission electron microscope 
(TEM)~\cite{Williams-1996-TTE}, scanning electron microscopev~\cite{Seiler-1983-SEE}. 
Scanning tunneling microscope magnification up to 300 million times~\cite{Tersoff-1985-TOT}. 
Furthermore, there are other types of microscopes, such as atomic force 
microscope~\cite{Binnig-1986-AFM}, Raman microscope and cryo-electron 
microscopy~\cite{Duncan-1982-SCA,Adrian-1984-CMO}.

Microscopic image analysis has a wide range of application scenarios, such as 
microorganism image analysis~\cite{Li-2020-ASM,Li-2019-ASF,zhang-2021l-LAN}, histopathological 
image analysis \cite{Li-2021-ACR,Zhou-2020-ACR}, cytopathological image 
analysis~\cite{Rahaman-2020-ASF,Hore-2015-FCO}, metal structure analysis~\cite{Oien-2014-DSA}, 
rock structure analysis~\cite{Clelland-1991-ARC}, soil structure analysis~\cite{Pagliai-2002-IAA}, 
material structure analysis and image analysis in plant 
pathology~\cite{Abell-1999-MIP,Nilsson-1995-RSA}.

\subsection{Background Knowledge of Intelligent Microscopic Image Analysis }
Manual operations of microscopic image analysis has some limitations, including: 
\begin{itemize}
\item In the case of big data, it takes a long time.
\item Operator has a heavy workload. 
\item Analysis of experimenters is easy to be subjective. 
\item Poor quantification. 
\end{itemize} 
Therefore, upcoming technologies should be introduced to this field to improve the level 
of image analysis.

As mentioned above, efficient artificial intelligence technology is introduced to the 
field of microscopic image analysis, which can effectively solve the above problems. 
First of all, computer efficiency is high and big data problems are easy to solve. 
Secondly, computer assistance can efficiently reduce the amount of hardwork. The third point, 
the computer is more objective. Finally, computer algorithms can quantitatively output 
the numerical results~\cite{Nilsson-2014-POA}. Especially, deep learning is the game changer 
and effective machine learning technology in the field of artificial intelligence in recent 
years, which has a high accuracy (ACC) rate. With the increasing amount of training data, 
the ACC rate is higher~\cite{Shen-2017-DLI}. Deep learning~\cite{litjens-2017-ASO} has 
a strong learning ability, wide coverage, adaptability and good portability. In different 
deep learning methods, CNN has the characteristics of non-contact and high precision for 
image recognition, classification and other operations. CNN is extremely applicable to this 
non-contact method in the process of image segmentation, detection, identification and 
classification, which can directly take image data as input~\cite{Tajbakhsh-2016-CNN}. 
Among all well-known CNN methods, U-Net~\cite{Ronneberger-2015-UCN} is clearly the most 
successful model for microscopic image segmentation, which does not require too many training 
data to obtain better segmentation results. Furthermore, the training and test time of U-Net 
is shot, which supports a practical possibility for some real tasks, such as histopathology 
image analysis, microorganism image analysis and cell image analysis.

\subsection{Typical U-Net Architecture}
\begin{figure}[htbp!] 
  \centerline{\includegraphics[width=0.8\textwidth]{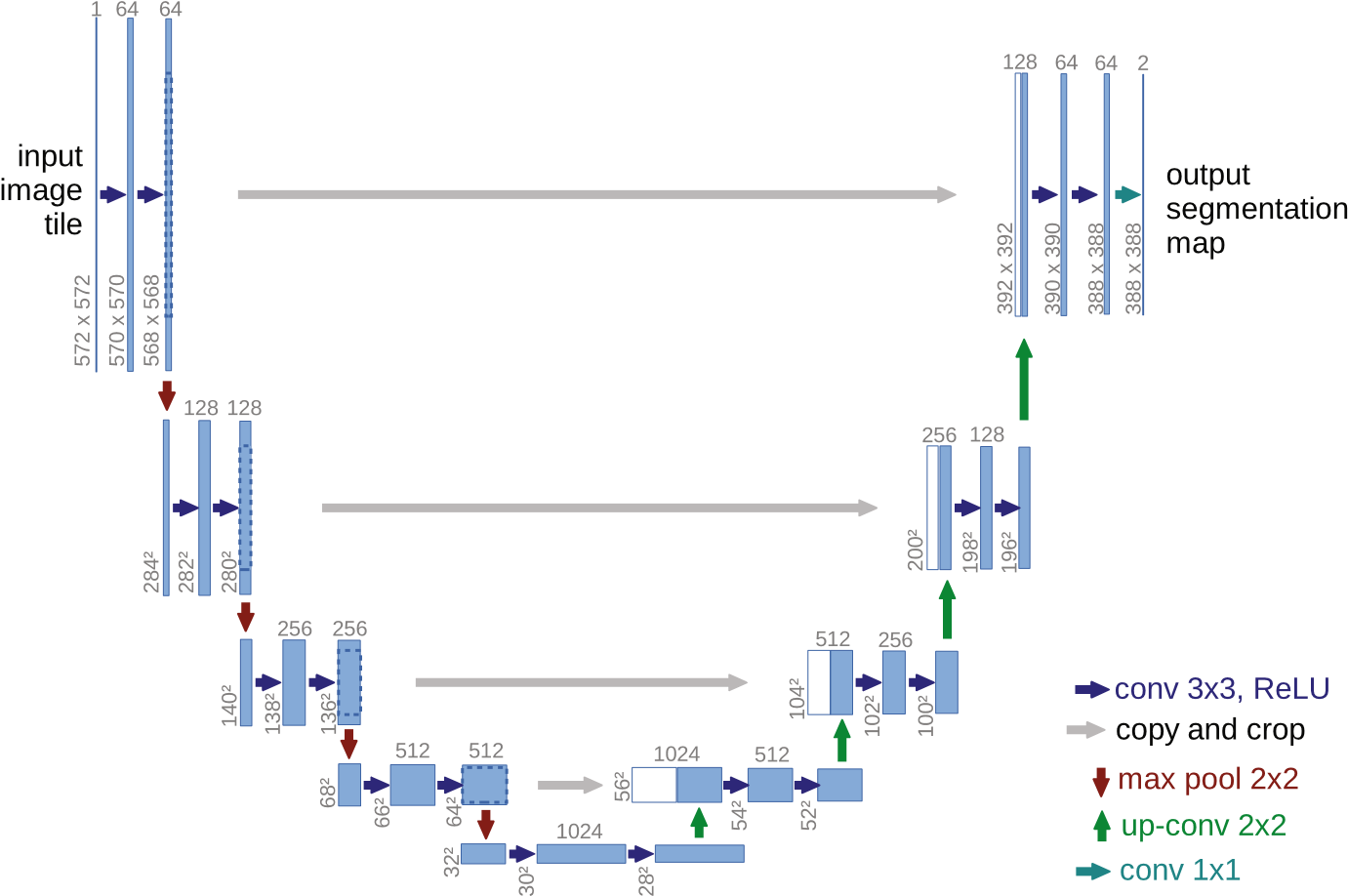}}
  \caption{The network structure of a typical U-Net~\cite{Ronneberger-2015-UCN}.} 
  \label{Fig.1}
\end{figure}

U-Net is originated from Fully Convolutional Networks(FCN), which is a semantic segmentation 
network \cite{Long-2015-FCN,Ronneberger-2015-UCN}. This network is named U-Net because of the 
U-shaped network structure, which is very suitable for medical image segmentation. The network 
structure is shown in Fig.~\ref{Fig.1}. Structure of U-Net is symmetrical. The left side is 
the encoder which can extract the input features, the right side is the decoder, which can 
output the encoded features as a picture. Red square represents down-sampling, green square 
represents up-sampling and conv $1 \times 1$ represents the convolution operation with 
$1 \times 1$ as the core network. The networks on the left are the traditional convolutional 
layer and pooling layer. The $3 \times 3$ convolution and rectified linear unit (ReLU) 
activation function are performed twice and there is a $2 \times 2$  maximum pooling. 
The process is repeated four times and the filter is doubled at each stage. The right part 
uses the $2 \times 2$ transposed convolution to upsampling. After performing the $3 \times $3 
convolution and ReLU activation function twice, it is also executed four times in a loop and 
the filter is reduced by half after each upsampling. Finally, add a $1 \times 1$ convolution 
plus SIGMOID activation function to get the result. In addition, every time downsampling is 
performed when preparing for pooling, it will be fused to the feature map after transposed 
convolution, so that the output on the encoder can be directly connected to the decoder to 
continue propagation. It can be seen that the network is not fully connected. This is also 
an end-to-end image, that is, the input is an image and the output is also an image. 

\subsection{Motivation of This Survey Paper }
Because of the efficiency and popularity of U-Net in microscopic image segmentation, many 
researchers choose to use it. Currently, there are some review articles involving U-Net, but 
there is no survey that focuses on U-Net in  microscopic image analysis. Hence, we decide 
to prepare this suvery paper to organize and summarize lated works for future work as a reference.

\cite{Taghanaki-2020-DSS} introduces the development of segmentation neural network 
from the perspective of deep semantic segmentation of natural and medical images, including the 
development of U-Net and the variant of U-Net. There are 163 references in this paper, but only 
16 are about microscopic image analysis. In the work of~\cite{Du-2020-MIS}, a review of U-Net 
is given from the use of U-Net in different application scenarios. There are 63 references 
in this work, but only eight references about microscopic image analysis are discussed. 
Compared with the above published reviews, we summarize 158 related works from 2015 to 2021, 
involving microorganism images, histopathological images, cytopathological images, rock 
microscopic images, metal microscopic images, plant microscopic images and material microscopic 
images. Our survey paper can provide services for the following two groups of people: 
(1) People who focus on microscopic image analysis; (2) People who do related to deep learning 
research and development, such as U-Net.

\subsection{Structure of This Paper}
The general structure of this paper is as follows: The first section summarizes 
the background of microscopic images, development status, motivation for this work and the 
structure of the paper. Section two describes the application scenarios of the unmodified U-Net. 
The third section writes the deformation method of U-Net under the primary changes and the 
dataset and application scenarios that it is good at segmenting after deformation. The fourth 
section is about some advanced and complex changes with Residuals in U-Net. This section 
contains many different types of changes.

\section{ Undeformed Classic U-Net } 
This section discusses  about the unmodified U-Net in microscopic image segmentation 
and summarize the literature.

\subsection{Applications in Cytology}
\begin{figure}[htbp!]
 \centerline{ \includegraphics[width=0.9\textwidth]{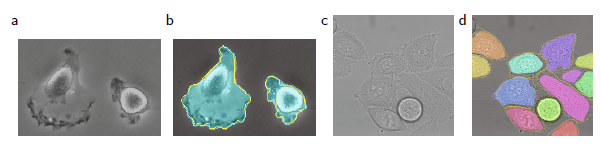}}
\caption{Results of the method proposed in~\cite{Ronneberger-2015-UCN} Fig.4. 
(a) An original image in ``PhC-U373'' dataset. (b) Segmentation result of (a). 
(c) An original image in ``DIC-HeLa'' dataset. (d) segmentation result of (c).
}
\label{Fig.11}
\end{figure}

In~\cite{Ronneberger-2015-UCN}, U-Net is proposed to solve a problem of microscopic image 
segmentation. In this work, original images ($512 \times 512$ pixels) are resized into 
$572 \times 572$ pixels to fit the input scale of U-Net. The method is applied to three datasets.    
Neuron structure has 30 pieces of training data. First, ``PhC-U373'' dataset has 35 training 
images with a small number of annotations. Second, ``DIC-HeLa'' dataset has 20 training images, 
some of them are annotated.  At last, ``PhC-U373'' and ``DIC-HeLa'' obtain intersection over 
union (IOU) of $92.03\%$ and $77.56\%$, respectively. Segmentation results are show in 
Fig.~\ref{Fig.11}. In Sec.~\ref{Sec:4.1.2}, we will discuss about 3D U-Net. 

In~\cite{Colonna-2018-SOC}, undeformed U-Net can accurately segment corneal nerves. A dataset 
contains 30 people is tested, of which 10 are healthy  and 20 are sick. This dataset is 
adjusted appropriately: first, outermost periphery (10 pixels) of the image to be analyzed 
is cropped. In addition each image is reduced to 0.7 times of the original size. Experimental 
result shows that  SE reaches $97.2\%$.

In~\cite{Seong-2019-AIO}, an original U-Net is used to automatically segment nerve cells. 
In this paper, a classification task of excitatory, inhibitory neurons and glia cells is 
also performed. The original U-Net is trained on 126 images containing 5000 cells and an 
experimental result of an ACC of $93.2\%$ is obtained.

Since the change of density of corneal endothelial cells (CEC) can monitor corneal diseases. 
In~\cite{Daniel-2019-ASO}, a baseline U-Net is used to automatically segment corneal endothelium 
in a large set of ``real-world'' specular microscopy images. Compared with the baseline 
U-Net~\cite{Ronneberger-2015-UCN}, neural network structure used for segmentation that has no 
structural changes. A mirror microscope database containing 16000 images (from corneal 
consultation service) is selected to evaluate this method and 158 training images are randomly 
selected as the training set. Finally, the experimental results with a CEC recall (RE) of $34\%$ 
and an ACC of $84\%$ are obtained.

\subsection{Applications on Microorganisms}
In~\cite{Nunez-2020-ASS}, Tuberculosis (TB) cords are segmented by U-Net, taking into account 
the ACC of U-Net segmentation of TB cords. This method is also easy to operate when small 
reserve of expertise. Datasets are obtained through image reconstruction, a total number 300 
images, of which 120 sub-images form the training set and 30 sub-images form the test set. 
Then, U-Net processes and segments sub-images. Finally, this method connects each sub-image 
to TB cords for the full images and the segmentation is completed. Finally, this work 
achieves an IoU of $88\%$ and an ACC of $92.01\%$.

In~\cite{Ojeda-2020-CNN}, in order to reduce the time and physical waste of the operator, 
U-Net is used to segment the parasites in the blood to implement an automated system. This 
method is tested on a private dataset composed of 974  images. In the  dataset, 600 images 
form the training set and 200 images form the testing set. The result shows that this method 
obtains an ACC of $63.04\%$ and a Dice similariy coefficient (DICE) score of $68.25\%$.

\subsection{Other Applications}
\begin{figure}[htbp!] 
  \centerline{\includegraphics[width=0.5\textwidth]{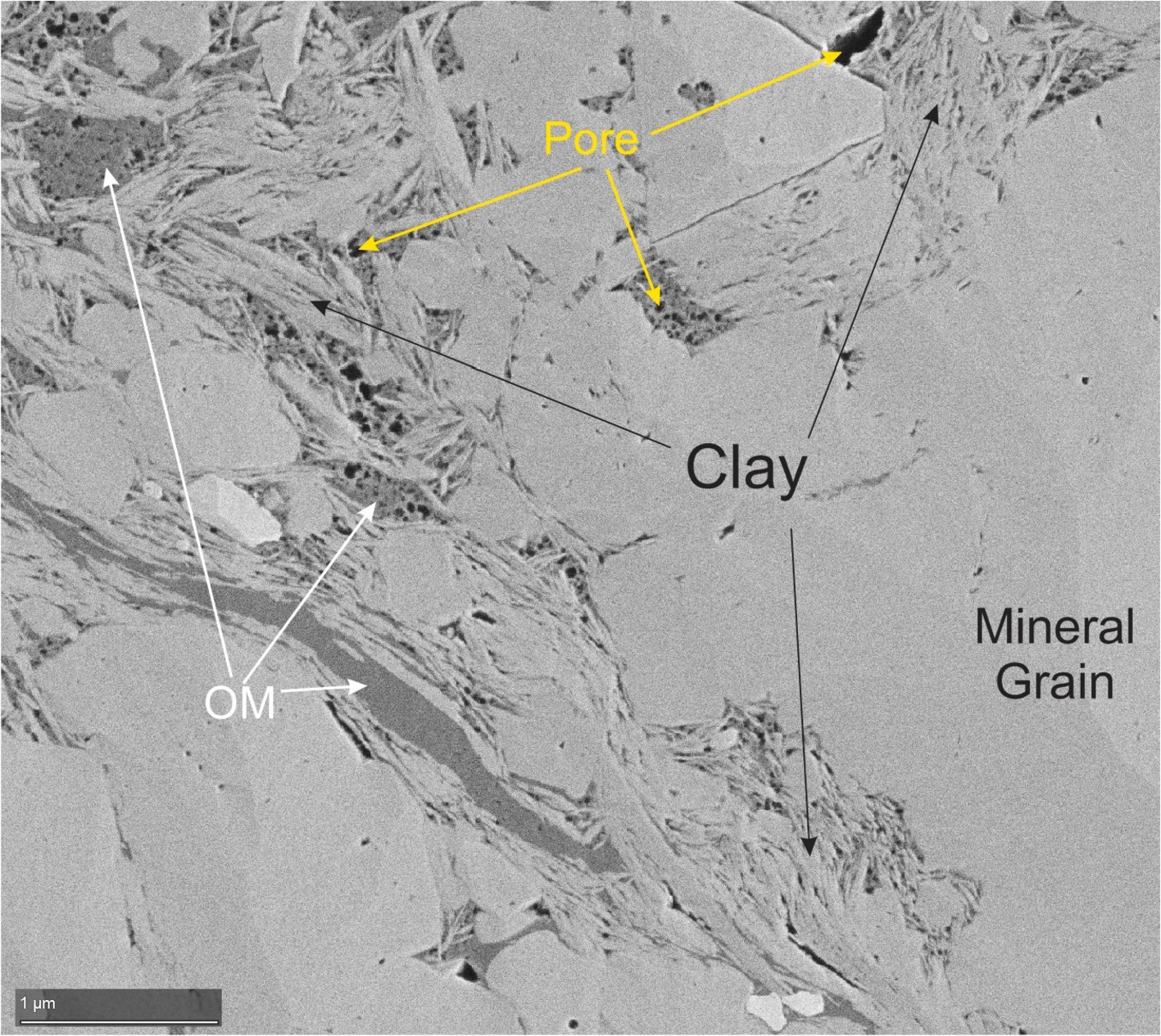}}
\caption{An example of electron SEM image in~\cite{Chen-2020-DLB} (Fig.1).}
\label{Fig. 3}
\end{figure}

In~\cite{Chen-2020-DLB}, in order to evaluate the geological characteristics of the rock 
samples, U-Net is used to segment the scanning electron microscope (SEM) images of it (as 
shown in Fig.~\ref{Fig. 3}). The dataset contains 8000 rock slice images, of which $80\%$ 
are used for training and $20\%$ for testing. The experiment result shows that the highest 
IOU reaches $93.2\%$. It can be seen that when processing texture features, U-Net has a 
good segmentation performance.

In~\cite{Oktay-2019-ADL}, a new method of detecting nanoparticles is proposed, in which 
U-Net is used to segment nanostructures. The dataset has eight TEM images synthesized by 
Fe$_{3}$O$_{4}$  nano-particles and nine images of Fe$_{3}$O$_{4}$  nano-particles smeared 
with silica, a total of 17 images. The experiment obtains an Acc higher than $90\%$.

\begin{figure}[htpb!]  
\centerline{\includegraphics[width=0.8\textwidth]{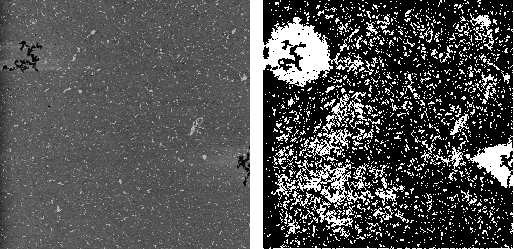}}
\caption{An example of a multilayer image segmented using a U-Net model. 
This figure corresponds to Fig.9 in~\cite{Farley-2020-ITS}.}
\label{Fig. 10}
\end{figure}

In~\cite{Farley-2020-ITS}, because manual segmentation of nanostructured surface images is 
time-consuming and requires high professionalism. Therefore, it is necessary to develop an 
automated method for segmenting nanoparticles based on U-Net. A method based on U-Net is 
better than the traditional automation method. The experimental dataset contains 728 images 
of gold nanoparticles generated by the atomic force microscope (AFM) in the absence of moisture, 
of which $75\%$ are used as the training set and $25\%$ are used as the test set. When banding, 
the average rate of U-Net pixel change is $12.2\%$. The result of segmentation with U-Net is 
shown in Fig.~\ref{Fig. 10}. 

Skin healthy condition is evaluated through detection and analysis of blood vessel structure. 
In~\cite{Jaworek-2018-ADL}, a fully automatic method using a baseline U-Net is developed to 
segment the blood vessel structure in dermoscopy color images. This work utilizes a dataset 
from the University of Naples, Italy and the University of Graz, Austria, which contains 74 
images of different types of blood vessel patterns in total. In the dataset, $80\%$ of the 
samples are used for training and the remaining $20\%$ are used for testing. The final 
experimental results are: a sensitivity (SE) of $85\%$ and a specificity (SP) of $81\%$.

Many diseases can be diagnosed by retinal blood vessels, in~\cite{Meyer-2017-ADN}, the original 
U-Net is used to segment Scanning Laser Ophthalmoscopy (SLO) (introduced by~\cite{Webb-1980-FST})  
images. SLO assists blood vessels to be extracted features better. This proposed method is 
evaluated by IOSTAR~\cite{Zhang-2016-RRV}. In IOSTAR, 20 images form the training set and 
10 images form the test set. The proposed method obtains an experimental result of an 
SE of $80.38\%$, a SP of $98.01\%$, an ACC of $96.95\%$ and an 
area under curve (AUC) of $97.71\%$, respectively.

Since a method of manually obtaining tissue sections can easily introduce errors (such as 
deformation, tissue fracture). Therefore, the baseline U-Net~\cite{Swiderska-2018-DLF} is 
developed to segment the undesired areas in whole slide images (WSIs) to reduce errors. 
This proposed method can be used in the preprocessing step of WSIs automated analysis. The 
method may prevent the damaged area from being used. Meanwhile, the classification task is 
carried out by the baseline U-Net. Original WSIs are classified as Damaged and non-Damaged classes. 
This work utilizes a dataset from the archives of the Department of Pathology at the 
Military Institute of Medicine in Warsaw, Poland.  The dataset contains 34 brain tissue 
cohorts corresponding to brain tumor areas (meningiomas and oligodendrogliomas) in total. 
However among the dataset, 10 WSIs for training and 24 WSIs be used in testing. Regarding 
Ki-67 brain tumor specimens segmentation, an experimental result is as follows: SE = 0.83, 
SP = 0.92, precision (PR) = 0.80, ACC = 0.90 and IoU = 0.69.

\subsection{Summary}
From the above analysis and review we can reach to a conclusion that the significant 
segmentation performance of U-Net on microscopic images, the original U-Net without any changes 
is often used to perform segmentation tasks and achieve good results. It shows that U-Net has 
strong versatility. U-Net also has wealthy application scenarios, such as cytology, histopathology,  
microorganism and nanoparticle image analysis. From birth of U-Net in 2015 to 2021, the original 
U-Net is used in a total number of papers that are listed above. With the development over the 
time, more and more variant network structures based on U-Net appears. In the subsequent 
analysis of the deformed structure section, the specific U-Net based on deformed network 
structure introduced. Tab.~\ref{Table. 1} is a summary of the evaluation indicators and results 
in the paper using the original U-Net.

\begin{table}[htbp!]
\caption{Summary of the original U-Net for segmentation tasks. 
The second column ``Detail'' shows the application object.}
\label{Table. 1}
\small 

\renewcommand\arraystretch{2}
\setlength{\tabcolsep}{1pt}          
\resizebox{\textwidth}{40mm}{
\newcommand{\tabincell}[2]
{\begin{tabular}{@{}#1@{}}#2\end{tabular}}

\begin{tabular}{|c|c|c|c|c|c|c|c|}
\hline
Aim                            & Detial                                                                                             & Year & Reference & Team                  & Data Information                                                                                                                                          & CNN type & Evaluation                                                                                                          \\ \hline
\multirow{12}{*}{ \tabincell{c}{Segmentation}} &  \tabincell{c}{Glioblastoma-astrocytoma\\      U373 cells}                 & 2015 & \cite{Ronneberger-2015-UCN}        & O-Ronneberger         & 35  annotated images                                                                                                                                      & U-Net    & IoU = 92.03\%                                                                                                     \\ \cline{2-8} 
                               & corneal nerves                                                                                     & 2018 & \cite{Colonna-2018-SOC}        & A.Colonna             &  \tabincell{c}{30 images\\      10 healthy images and 20 pathological images\\       image is reduced to 0.7 times (256 X   256)} & U-Net    & SE = 97.2\%                                                                                                       \\ \cline{2-8} 
                               & nuclei                                                                                             & 2019 & \cite{Seong-2019-AIO}        & S.Seong               &  \tabincell{c}{              126 images\\      containing about 5000 cells}                                                                     & U-Net    & ACC =  93.2\%                                                                                                     \\ \cline{2-8} 
                               & corneal endothelium                                                                                & 2019 & \cite{Daniel-2019-ASO}        & M.C.Daniel            & 158  training  images                                                                                                                                     & U-Net    & RE = 34\%,    PR  = 84\%                                                                               \\ \cline{2-8} 
                               & TB cords                                                                                           & 2020 & \cite{Nunez-2020-ASS}        & L.Ballan              &  \tabincell{c}{300 images\\      120 sub-images for  training, 30   sub-images for test}                                          & U-Net    & IoU = 88\%,  ACC = 92.01\%                                                                                        \\ \cline{2-8} 
                               & the T. cruzi parasite                                                                              & 2020 & \cite{Ojeda-2020-CNN}        & A.Ojeda-Pat           &  \tabincell{c}{974 images\\      600 images for training, 200 images for validation}                                              & U-Net    & ACC = 63.04\%,  DICE =   68.25\%                                                                                  \\ \cline{2-8} 
                               & rock slice    SEM images                                                                           & 2020 & \cite{Chen-2020-DLB}        & Z.Chen                &  \tabincell{c}{8000  images\\      6400 for train  training, 1600 for   validation}                                               & U-Net    & IoU = 91.7\%                                                                                                      \\ \cline{2-8} 
                               & nano-particles                                                                                     & 2019 & \cite{Oktay-2019-ADL}        & A.B.Oktay             & 17 images                                                                                                                                                 & U-Net    & SE = 78.59\%,  ACC = 96.59\%                                                                                    \\ \cline{2-8} 
                               & nanostructured surfaces                                                                            & 2020 & \cite{Farley-2020-ITS}        & S.Farley              &  \tabincell{c}{728 images\\      546 for training, 182 for test}                                                                  & U-Net    & average proportion of pixels changed pixels changed = 12.2\%                                                      \\ \cline{2-8} 
                               &  \tabincell{c}{blood   vessel structure \\in dermoscopy color      images} & 2018 & \cite{Jaworek-2018-ADL}        & J.Jaworek-Korjakowska &  \tabincell{c}{74 images\\      \%80 for training, \%20 for test}                                                                 & U-Net    & DICE = 0.84,  SE =   0.85,  SP = 0.81                                                                              \\ \cline{2-8} 
                               & retinal blood vessels                                                                             & 2017 & \cite{Meyer-2017-ADN}        & M.I.Meyer             &  \tabincell{c}{30 SLO images\\      20 images\\      20 for training, 10  for test}                                               & U-Net    &  \tabincell{c}{SE = 0.8038, SP = 0.9801 \\       ACC = 96.95\%,  AUC = 0.9771}             \\ \cline{2-8} 
                               &  \tabincell{c}{empty   \\      areas in  brain tissue  WSIs}         & 2018 & \cite{Swiderska-2018-DLF}        & Z.Swiderska-Chadaj    &  \tabincell{c}{34 WSIs\\      Ki-67 staining\\      10 for training, 24  for test}                                                & U-Net    &  \tabincell{c}{SE = 0.83,  SP = 0.92,    PR = 0.80\\      ACC = 90\%,  IoU = 0.69} \\ \hline
\end{tabular}}
\end{table}

\section{Simple and Low-level Deformable U-Net}
From the above section we can see: in most cases, the original U-Net segmentation without any 
changes performed well. However, firstly, U-Net published in 2015. In recent years, a large 
number of new network structure changes  appear, such as: many new network modules are proposed 
and new ideas in skip connection are proposed. Secondly, U-Net can not fit all datasets perfectly. 
In many cases, the segmentation performance is not ideal, therefore, it is appropriate and 
simple to change the U-Net for better performance over the segmentation tasks. This section 
introduces some simple and conventional methods to change original implementation of U-Net 
structure.

\subsection{Redesigned Convolution}
\subsubsection{3D Convolution}
\label{Sec:4.1.2}
In~\cite{Cciccek-2016-3UN}, a 3D U-Net that replaces 2D convolution with 3D convolution is 
creatively proposed to achieve accurate semi-automatic segmentation of 3D datasets. Compared 
with baseline U-Net, 3D U-Net replaces all 2D operations in the paper~\cite{Ronneberger-2015-UCN} 
with 3D operations and adds Batch Normalization (BN) before each ReLU (similar addition BN’s 
approach is also reflected in the paper~\cite{Fang-2019-NSB}). A dataset with 77 manually 
annotated Xenopus kidney slices under confocal microscopic is used in the experiment. Under 
the dataset, the proposed 3D U-Net obtains an IoU of 0.863 in the semi-automated test 
experiments of three-fold cross validation.

\begin{figure}[htbp!]  
\centerline{\includegraphics[width=0.75\textwidth]{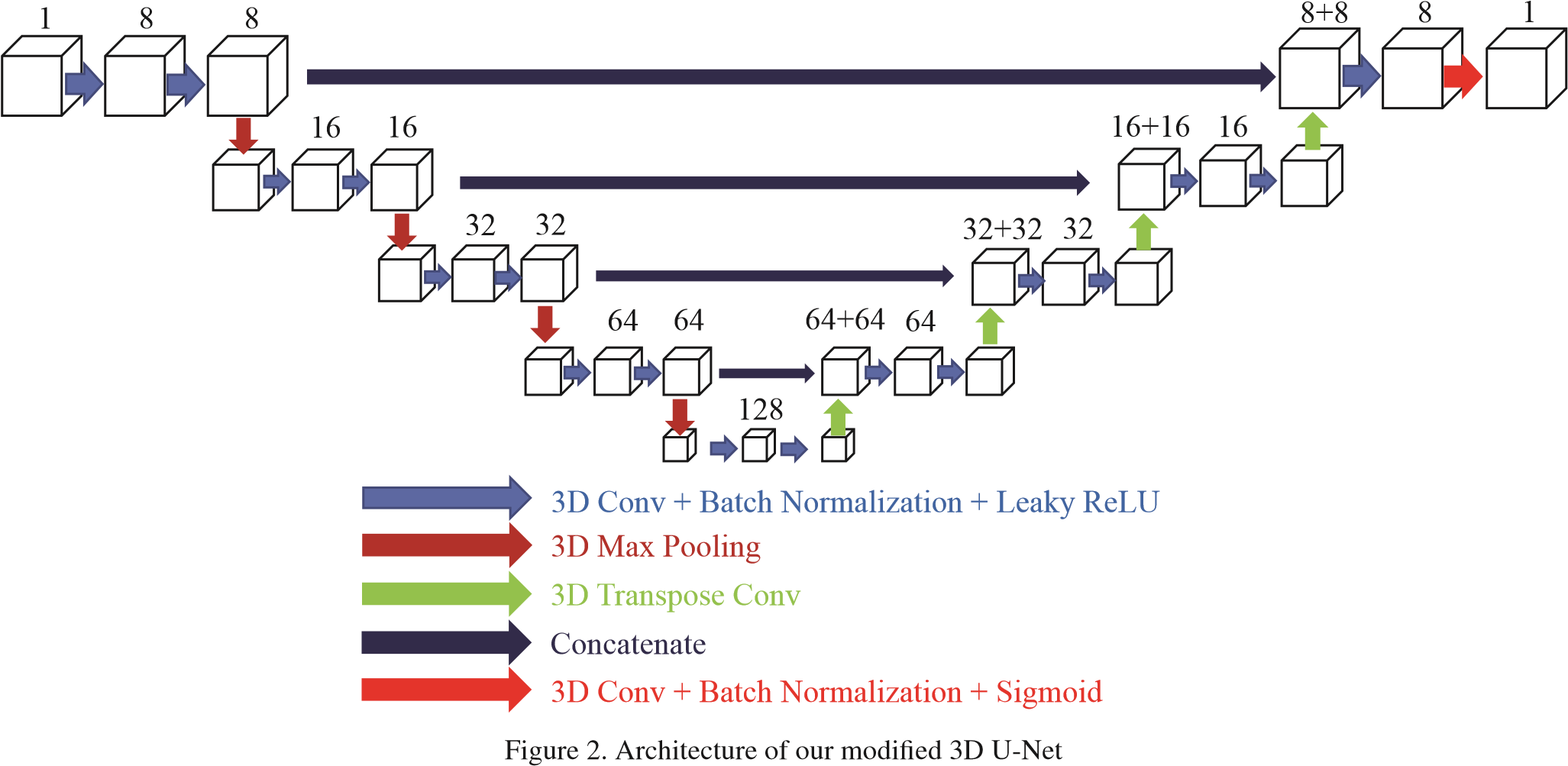}}
\caption{The network architecture of 3D U-Net (corresponding to Fig.2 in~\cite{Fu-2018-TDF}). }
\label{Fig.6}
\end{figure}

In~\cite{Fu-2018-TDF}, a 3D U-Net composed of 3D convolution combined with spatially 
constrained cycle-consistent adversarial networks is proposed. It solves the problem of high 
learning rate for manually labeling 3D datasets. The structure of this 3D U-Net is shown in 
Fig.~\ref{Fig.6}. First, a subvolume of the original image volumes train spatially constrained 
CycleGAN (SpCycleGAN). Then, the above operations generate 3D synthetic data to evaluate the 
3D U-Net. Finally, the proposed method obtains an ACC of $95.56\%$. 

\begin{figure}[H]  
 \centerline{\includegraphics[width=0.9\textwidth]{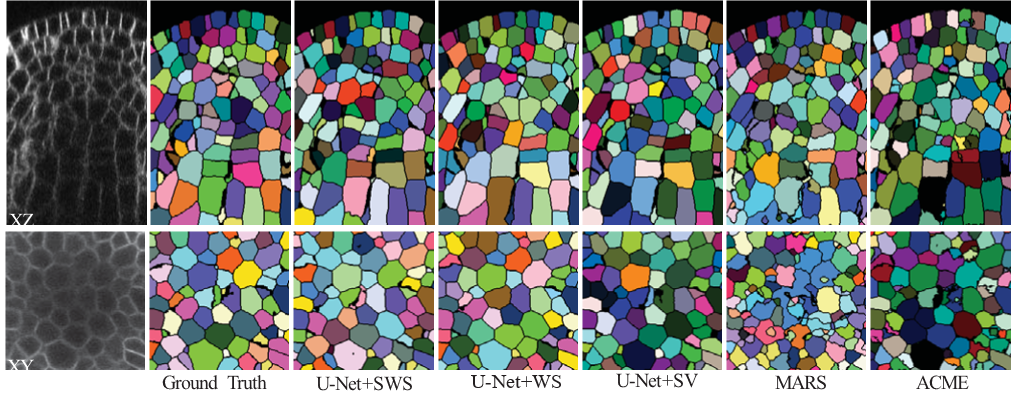}}
\caption{A comparison of the image segmentation reults of the proposed method 
in~\cite{Eschweiler-2019-CBP} (Fig.3), where the top row shows XZ-plane and the bottom row 
shows XY-plane.
}
\label{Fig.7}
\end{figure}

In~\cite{Eschweiler-2019-CBP}, a 3D U-Net  combined with a seeded watershed approach (SWS) is 
proposed. It solves a problem of the large amount of data in the 3D microscopic image of the 
cell membrane, which is not easy to segment. This method is tested on a training set (the 
training set comes from~\cite{Willis-2016-CSA}) composed of 109296 Arabidopsis thaliana cells 
and a validation set composed of 972 single cells, which manually annotate 3D images. At the 
same time, 3D U-Net $+$ watershed algorithm (WS), 3D U-Net $+$ supervoxel merging approach (SV), 
multi-angle image acquisition, three-dimensional reconstruction and cell segmentation 
(MARS)~\cite{Fernandez-2010-IPG}, automated cell morphology extractor for comprehensive 
reconstruction of cell membranes (ACME)~\cite{Mosaliganti-2012-AAC} are also tested by the 
dataset. The experimental results show that among these methods, U-Net $+$ SWS performs best 
and it obtains a JI of 0.870 and a DICE of 0.931. Fig.~\ref{Fig.7} is the segmentation result 
of the method mentioned in~\cite{Eschweiler-2019-CBP}.

In~\cite{Heinrich-2018-SCS}, a 3D U-Net is proposed to fit the non-isotropic nature of 
serial section Transmission Electron Microscopy (ssTEM) data, the sparsity of synapses improves 
the performance of segmentation and detection of insect nervous system. Challenge on Circuit 
Reconstruction from Electron Microscopy Images (CREMI) datasets from Medical Image Computing 
and Computer Assisted Intervention Society (MICCAI) is used in the experiment. It contains 6 
volumes of nerve tissue under an electron microscope, of which $75\%$ are used in the training 
set and $25\%$ are used in the validation set. The proposed 3D U-Net obtains the experimental 
results of CREMI score with an average of $50\%$.

In~\cite{Wang-2019-SNS}, because of the complex structure of neurons and poor imaging quality 
in some cases, a teacher-student learning framework based on 3D U-Net~\cite{Cciccek-2016-3UN} 
is proposed to segment neurons to obtain higher ACC and efficiency. 
Like~\cite{Cciccek-2016-3UN,Fu-2018-TDF,Eschweiler-2019-CBP,Heinrich-2018-SCS}, 
the teacher-student learning framework uses 3D convolution as the basic unit. 
Unlike~\cite{Cciccek-2016-3UN,Fu-2018-TDF,Eschweiler-2019-CBP,Heinrich-2018-SCS}, 
first, the teacher-student network is divided into 2 parts: Teacher network and student network. 
Second, Residual modules (the specific usage of Residual modules is detailed in Sec.~\ref{Sec.4}) 
are added to the teacher-student network. A data is obtained from the Janelia dataset from 
the BigNeuron project (contains 42 images of adult Drosophila nervous system, 38 images for 
training and 4 images for testing). A PR-RE curves in~\cite{Wang-2019-SNS} shows 
that the teacher-student network can obtain more accurate segmentation performance.

\subsubsection{Other Convolution}
\begin{figure}[htbp!]  
 \centerline{\includegraphics[width=0.65\textwidth]{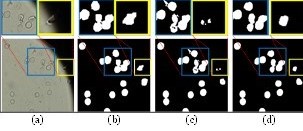}}
\caption{The cell segmentation results in~\cite{Zhang-2017-ISA} (Fig.3).
(a) Raw images; (b) Ground truth images; (c) Segmentation result of U-Net;
(d) segmentation result of deformable U-Net.}
\label{Fig.4}
\end{figure}

In~\cite{Zhang-2017-ISA}, deformable U-Net with variable convolution is used to segment and 
classify sickle cell disease (SCD) cells. Similarly, \cite{Zhang-2018-RSS} is the study of 
SCD cell segmentation from the same team. Compared with U-Net, in addition to variable 
convolution, deformable U-Net has only three deformable convolution blocks in the encoding 
and decoding parts respectively. A dataset with 128 SCD cells obtained from University of 
Pittsburgh Medical Center (UPMC) is used in the experiment. The deformable U-Net is tested 
under the dataset. It obtains a segmentation ACC of $97.8\%$ and a classification ACC of 
$82.7\%$. The segmentation results of proposed method in  Fig.~\ref{Fig.4}.

\begin{figure}[htbp!]  
 \centerline{\includegraphics[width=0.6\textwidth]{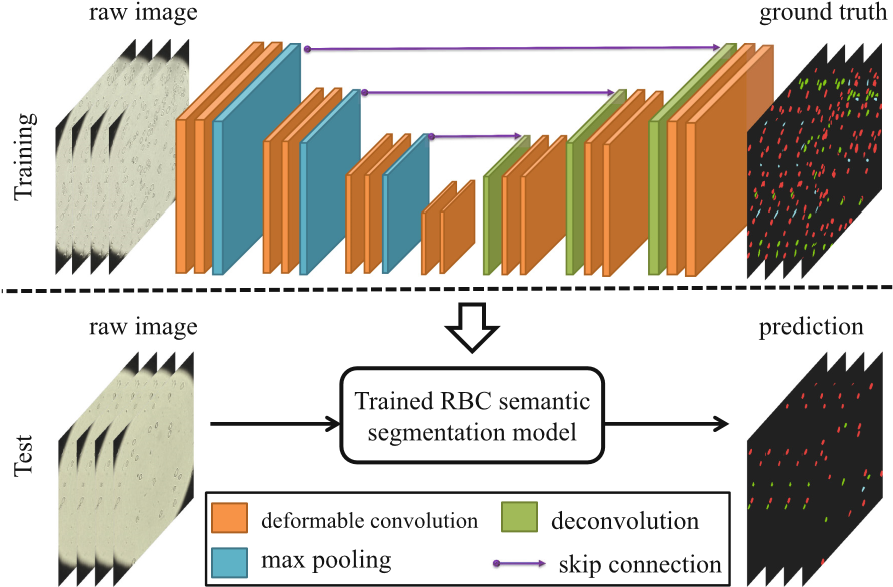}}
\caption{The network architecture of the deformable U-Net in~\cite{Zhang-2018-RSS} (Fig.2).}
\label{Fig.5}
\end{figure}

In~\cite{Zhang-2018-RSS}, in order to solve the inaccurate segmentation and classification of 
SCD cells are caused by the change of cell shape and the image blur caused by noise and 
artifacts, deformable U-Net with variability convolution is proposed. The structure of the 
deformable U-Net is shown in Fig.~\ref{Fig.5}, which is consistent with the network structure 
of~\cite{Zhang-2017-ISA}. The ordinary convolution becomes the deformable convolution and the 
convolution block is reduced. In contrast to~\cite{Zhang-2017-ISA}, the dataset in this paper is 
a public dataset of red blood cells (RBCs) of SCD patients, from~\cite{Xu-2017-ADC}, there are 
four different types of 266 original images. Under the dataset, they proposed new network is 
tested to achieve a segmentation ACC  of $99.12\%$ and an IoU of $44.15\%$.

In~\cite{Qin-2020-MFU}, a Match Feature U-Net used in the field of medical image dynamic 
reception is proposed to perform cell segmentation. 
Like~\cite{Zhang-2018-RSS,Zhang-2017-ISA,Cciccek-2016-3UN,Fu-2018-TDF,Heinrich-2018-SCS,Wang-2019-SNS}, 
Match Feature U-Net improves the ability to segment-specific or public datasets by changing 
basic convolution units. 
Unlike the above works that plain convolution becomes 3D convolution and deformable convolution, 
an adaptive receptive field mechanism is embedded in the Match Feature U-Net. The mechanism 
is introduced by adding a large number of Dynamic Convolution Units with Adaptive Receptive 
Field (ARF) convolution. Match Feature U-Net is evaluated by 670 cell nuclei (each original image 
is augmented to 16 times) from Data Science Bowl 2018, of which $80\%$ is used for 5-fold 
cross-validation and $20\%$ is used for testing. Finally, the Match Feature U-Net with Match 
operator obtains a MIoU of $91.83\%$ experimental results. Similar technical methods are not 
only used for cell segmentation. In~\cite{Rad-2018-BCC}, the U-Net-based method of changing 
convolution is applied to cell counting and centroid localization.

In~\cite{Rad-2020-TSI}, a variant U-Net composed of Inception modules (introduced 
in~\cite{Szegedy-2015-GDW}) with dilated convolution is proposed. The variant U-Net is proposed 
to obtain accurate segmentation of trophectoderm (TE) and to achieve an automatic evaluation 
of the quality of human embryos. It is different from deformed U-Nets formed by Inception 
modules, dilated convolutions with different dilation rates in the variant U-Net replaces 
plain convolutions. Datasets come from~\cite{Saeedi-2017-AIO} (contains 235 human blastocyst 
images) and a private dataset (including 592 human blastocyst images), of which $70\%$ are used 
as the train set and $30\%$ are used as the test set.

\subsection{Add convolution block, reduce convolution block}
In~\cite{Matuszewski-2018-MAT}, a deformed U-Net is used to segment images with minimal annotation. 
In addition to the reduction in the number of feature maps of the convolutional layer in the 
encoding and decoding part, the other parts of the deformed U-Net are consistent with the 
classical U-Net. One of the advantages of the deformed U-Net: the training parameters of the 
structure are reduced, which prevents overfitting. The deformed U-Net is tested by the Rift 
Valley virus dataset (there are 143 TEM images)~\cite{Kylberg-2012-SOV} and obtains a DICE of 
0.900 and an IoU of 0.831.

In~\cite{Mocan-2018-ADO}, in order to identify whether cells are normal or circulating tumor 
cells (CTCs), a modified U-Net is used to automatically segment the cells. The modified U-Net 
has three more layers in the encoding and decoding parts than the original U-Net structure. 
In addition, an additional $3 \times 3$ convolution and ReLU are added to each layer in the 
decoding part. Like~\cite{Xu-2019-UFR,Li-2017-NDL}, the modified U-Net from~\cite{Mocan-2018-ADO} 
has a change in the number of convolutional layers compared with the baseline U-Net.
Unlike~\cite{Xu-2019-UFR,Li-2017-NDL}, the modified U-Net from~\cite{Mocan-2018-ADO} segmentation 
of cell images in the blood instead of histopathological images. In terms of data, tumor cell 
datasets from the Oncology Institute of Cluj-Napoca are used. 120 image data expand into 56000 
images by generating  small patches, furthermore, $70\%$ of it is used as the training set and 
the rest is the test set. The final result shows that under their dataset, an ACC of $99.81\%$ 
is obtained.

By observing corneal endothelial cells, information about corneal health can be obtained in time. 
Because of the size of endothelial cells in the specular microscope image needs to be analyzed, 
a U-Net-based CNN is developed to segment endothelial cells~\cite{Fabijanska-2018-SOC}. 
Improvements compared with the baseline U-Net are described as follows: First, convolution blocks 
are reduced and the downsampling is reduced twice (the same operation is in~\cite{Zhang-2017-ISA}). 
Second, the number of feature vectors in each layer is $50\%$ from the original. A dataset employed 
in the experiment contains 30 images of the corneal endothelium. In the dataset, $50\%$ of the 
samples are used for training and the remaining $50\%$ are used for testing. A result of the 
experiment is that a DICE reaches $86\%$.

In~\cite{Xu-2019-UFR}, a new variant U-Net (named US-Net) is proposed for robust nuclei instance 
segmentation in histopathology images. 
Like~\cite{Li-2017-NDL}, US-Net has a post-processing part after the output layer.
Unlike~\cite{Li-2017-NDL}, US-Net combines a single shot multibox detector (SSD) to form a 
post-processing sub networks, rather than simply adding a post-processing layer.
A training dataset curated by the Segmentation of Nuclei in Images Contest (SNIC) and the 
Medical Image Computing and MICCAI. 32 patches from SNIC and 30 patches from MoNuSeg are 
pre-processed to obtain 878 patches, of which 650 patches are used for training and 228 patches 
are used for evaluation. an experimental result shows that the proposed US-Net performs better 
than many advanced nuclear detection and segmentation networks.

In~\cite{Li-2017-NDL}, noise-tolerant U-Net is proposed to fully automate the segmentation 
of histopathological images. Two differences between the noise-tolerant U-Net and the baseline 
U-Net are: noise-tolerant has two less convolutional blocks. However, it adds a noise-tolerant 
layer after a softmax output layer. Like~\cite{Xu-2019-UFR,Mocan-2018-ADO}, noise-tolerant 
U-Net changes the number of layers of convolution. Unlike~\cite{Xu-2019-UFR,Mocan-2018-ADO}, 
noise-tolerant U-Net reduces the convolutional layer. A dataset includes five groups of 
histopathological images of Duchenne Muscular Dystrophy (DMD). The first group contains 110 
images as a training set and the remaining four groups (100 images in each group) are used 
for validation and comparison.

However the original network requires more annotated images. Throughout the study 
of~\cite{kumar-2020-CSB}, a modified U-Net that only needs a small amount of annotated images and 
has a more appropriate amount of calculation is proposed. Compared with the classical U-Net, this 
modified U-Net has three main improvements: reducing the number of filters, adding a BN layer after 
the convolutional layer and adding rectified-Adaptive Moment Estimation (Adam). From ISBI cell 
tracking Challenge, 120 images (data augmentation from 30 images to 120 images by cutting and 
flipping) of the Drosophila first instar larva ventral nerve cord (VNC) are used to evaluate 
the modified U-Net. Experimental results show that an IoU of $92.54\%$ is obtained.

\subsection{Composite Structure Appearance Multiple U-Net Chains}
\begin{figure}[htbp!]  
 \centerline{\includegraphics[width=0.5\textwidth]{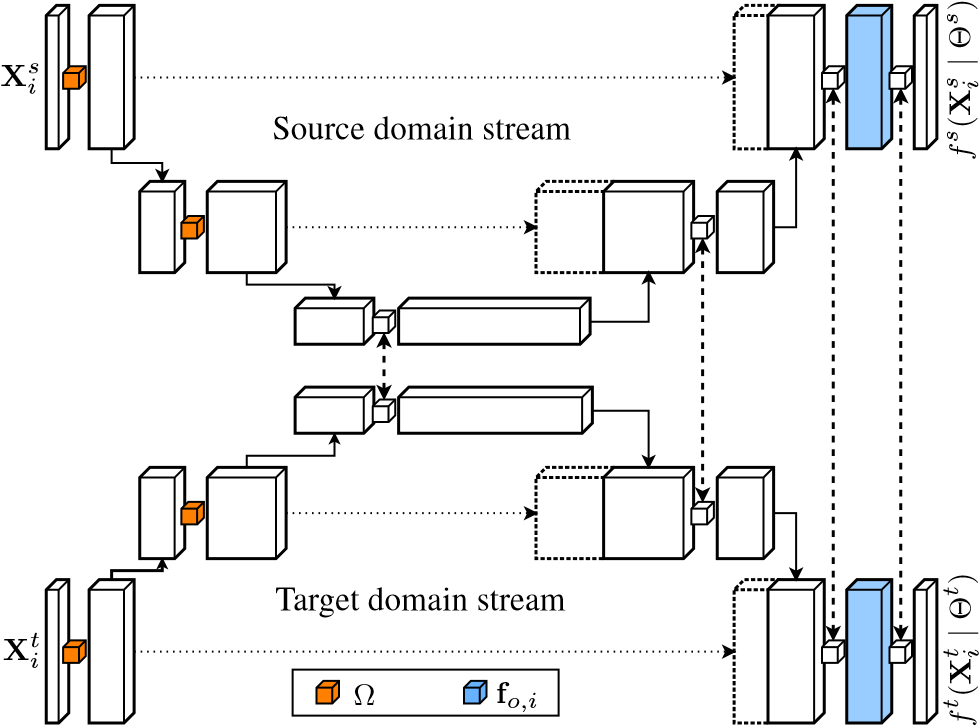}}
\caption{The network structure of two-stream U-Net in~\cite{Bermudez-2018-ADT} (Fig.2).}
\label{Fig. 51}
\end{figure}

In~\cite{Bermudez-2018-ADT}, since the mitochondria and synapses in the mouse brain under the 
electron microscope are not easily segmented, a two-stream  U-Net with two coupled U-Nets is 
proposed. Two-stream U-Net is different from the classic U-Net~\cite{Ronneberger-2015-UCN}. 
First of all, it consists of two symmetrically distributed U-Nets. Furthermore, one U-Net acts on 
the source domain and the another one acts on the target. Finally, the weight is shared by two 
U-Nets. Like~\cite{Jha-2020-DAD,Zhuang-2018-LMN}, two-stream U-Net consists of two 
similar U-Nets. Unlike~\cite{Jha-2020-DAD,Zhuang-2018-LMN}, two U-Nets of two-stream U-Net are 
trained by the source domain and the target domain, respectively. However, in~\cite{Jha-2020-DAD} 
and~\cite{Zhuang-2018-LMN}, the input of U-Net at the back is related to the output of U-Net 
at the front. TEM volumes of mouse somatosensory cortex and cerebellum are used to test the 
two-stream U-Net and a JI of 0.7230 is obtained. The architecture of two-stream U-Net is 
shown in Fig.~\ref{Fig. 51}.

In many segmentation tasks, the traditional encoder, decoder and skip connection structure 
cannot complete the task perfectly and there are few channels for information flow to circulate. 
In~\cite{Zhuang-2018-LMN}, LadderNet is proposed to solve the problem. The same point 
as~\cite{Bermudez-2018-ADT,Jha-2020-DAD}, the weights are shared between the two U-Nets. The 
difference from~\cite{Bermudez-2018-ADT,Jha-2020-DAD}, LadderNet shares the weights in the 
decoding part of the first U-Net and the encoding part of the second U-Net. But, two-stream 
U-Net shares weights in the decoding part of the two U-Nets. A dataset is obtained from one source: 
the CHASE DB1 dataset (contains 28 retinal images, of which $70\%$ is used for training and $30\%$ 
is used for testing). In the dataset, 0.8031 F1-score, 0.7978 SE, 0.9818 SP, 0.9656 ACC and 
0.9839 AUC are obtained.

\begin{figure}[htbp!]  
 \centerline{\includegraphics[width=0.75\textwidth]{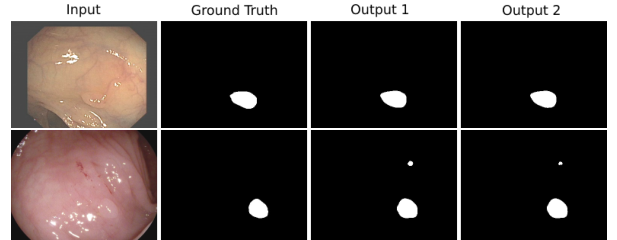}}
\caption{The segmentation results of DoubleU-Net in~\cite{Jha-2020-DAD} (Fig.3). }
\label{Fig. 52}
\end{figure}

In~\cite{Jha-2020-DAD}, a stacked double U-Net is proposed to obtain better segmentation ACC 
and is named DoubleU-Net. Like~\cite{Zhuang-2018-LMN,Bermudez-2018-ADT}, encoding and 
decoding ideas of U-Net are not changed. Unlike~\cite{Zhuang-2018-LMN,Bermudez-2018-ADT}, first, 
the encoding part of a U-Net in DoubleU-Net is replaced with VGG-19~\cite{Simonyan-2014-VDC}. 
Second, the information flows from the previous U-Net encoding part to the next decoding part. 
Third, DoubleU-Net is evaluated by the alike-microscopic dataset. DoubleU-Net is tested on the 
CVC-ClinicDB dataset~\cite{Bernal-2015-WMF} to obtain a DICE of 0.9239, a mean Intersection over 
Union (mIoU) of 0.8611, a 0.8457 RE and a 0.9592 PR. The segmentation results 
of DoubleU-Net are shown in Fig.~\ref{Fig. 52}.

The variant structure based on U-Net is not easy to segment adjacent cells, in~\cite{Torr-2020-DSO} 
DeepSplit is proposed to segment the cell contact areas. 
Like~\cite{Bermudez-2018-ADT,Zhuang-2018-LMN,Jha-2020-DAD}, DeepSplit has multiple U-Net chains 
with a composite structure appearance. Unlike~\cite{Bermudez-2018-ADT,Zhuang-2018-LMN,Jha-2020-DAD}, 
DeepSplit only has one encoder branch and two decoder branches. The second decoder branch is a 
separate branch, its main purpose is to segment the cell contact areas. This method is evaluated 
on an MCF-10a epithelial breast cells dataset (contains 50 manually annotated images, of which 
$80\%$ is used for training, of which $10\%$ is used for verification and of which $10\%$ is used 
for testing). Finally, a cell detection score (CDS) of 0.903 is obtained.

In~\cite{Bozkurt-2018-AMC}, a MUnet nested with three U-Nets is proposed, which is used to segment 
the image of morphological patterns of human skin under reflectance confocal microscopy (RCM). 
Because the traditional method is time-consuming, MUNet is used to assist in the diagnosis of this 
type of skin cancer. 56 RCM mosaics (46 mosaics are used for training and 10 mosaics are used for 
testing) annotated by experts with 6 types of labels are used to evaluate MUnet. Because three 
nested U-Net structures are designed, the segmentation operation of MUet can be performed at 
different resolutions. Finally, for the background segmentation under the dataset, MUNet obtains 
a $72.89\%$ SE, a $95.26\%$ SP, a $78.36\%$ DICE and a $84.71\%$ PR.

In~\cite{Zhao-2020-TUH}, a Triple U-Net is proposed for nuclear segmentation to avoid blurry 
tumor nucleus boundaries and overlapping tumor cells as much as possible.
Like~\cite{Bozkurt-2018-AMC}, the Triple U-Net is composed of three U-Net branches: a 
red-green-blue (RGB) branch, a Hematoxylin branch and a Segmentation branch. 
Unlike~\cite{Bozkurt-2018-AMC}, two branches of the Triple U-Net  are composed of Progressive 
Dense Feature Aggregation Module (PDFA) based on the densely connected block (be introduced 
by~\cite{Huang-2017-DCC}). This work utilizes the MoNuSeg dataset from 7 organs which contain 
30 images in total, 21000 nuclear boundaries are annotated. In the total dataset, 16 images are 
used for training and 14 images are used for testing. Comparison of Triple U-Net and existing 
nuclei segmentation models, the final Triple U-Net result is the best (0.837 DICE).

\subsection{Attention U-Net}
An attention mechanism is introduced into U-Net in~\cite{Oktay-2018-AUL} and applied to a CT 
dataset to highlight features. In~\cite{Lian-2018-AGU}, an ATTention U-Net (ATT-UNet) is proposed 
to segment an iris to solve a problem that a segmentation network is susceptible to irrelevant 
noise pixels outside the iris area. ATT-UNet enables a network to focus on a region of interest 
(ROI), avoid wasting time and over-computing features of irrelevant regions. The purpose of the 
attention mechanism is introduced to guide ATT-UNet to learn more features to separate the iris 
and non-iris pixels. An attention mask is generated to evaluate the most likely areas of the iris 
and a bounding box regression module is used to evaluate the coordinates. Furthermore, the 
attention mask guides ATT-UNet to segment the specific area. A dataset employed in the experiments 
comes from UBIRIS.v2~\cite{Proencca-2009-TUA}, of which 500 images are used as the training set 
and 500 images are used as the test set. In the end, an experiment result shows: an IoU of 
$91.37\%$ is obtained.

In~\cite{Lv-2020-AGU}, an attention guided U-Net with atrous convolution (AA-UNet) is proposed to 
segment retinal blood vessels. A precise segmentation of retinal blood vessels has an important 
auxiliary role in the diagnosis of diabetes, hypertension and other diseases. 
Like~\cite{Lian-2018-AGU}, an attention module is used to force the network to pay attention 
to ROI. Unlike~\cite{Lian-2018-AGU}, atrous convolution replaces ordinary convolution in the 
feature blocks, which is beneficial to increase the receptive field. AA-UNet is tested on three 
retinal vessels segmentation datasets (DRIVE~\cite{Bansal-2013-RVD}, STARE~\cite{Guo-2018-ARV}  
and CHASE$\_$DB1~\cite{Thangaraj-2018-RVS}). The DRIVE (Digital Retinal Images for Vessel 
Extraction) dataset contains 40 fundus images, of which 20 images are used for training and 
20 images are used for testing. Under the DRVIE dataset, AA-UNet obtains $95.58\%$ ACC, 
$82.16\%$ F1-scores, $95.68\%$ Jaccard similarity (JS) and $98.47\%$ AUC.

In~\cite{Mou-2019-CCA}, in order to segment curved structures (such as blood vessels), CS-Net 
is proposed to assist experts in diagnosing diseases. Channel attention block (CAB) and spatial 
attention block (SAB) with attention ideas are integrated into the baseline U-Net. Attention 
idea appears in the form of a module after the encoder. STARE is a fundus dataset used to evaluate 
this proposed method. An experimental result is obtained by CS-Net ($97.52\%$ ACC, $0.88.16\%$ SE, 
$98.40\%$ SP and $99.32\%$ AUC).

In~\cite{Li-1903-CSA}, a connection sensitive attention U-Net (CSAU) is proposed to segment 
retinal blood vessels. Like~\cite{Mou-2019-CCA,Lv-2020-AGU,Jiang-2020-multi}, CSAU is proposed 
to segment retinal blood vessels. Unlike~\cite{Mou-2019-CCA,Lv-2020-AGU,Jiang-2020-multi}, 
a connection sensitive loss is proposed and combines with attention gates. CSAU is trained on 
the DRIVE, STARE and HRF datasets, respectively. STARE contains 20 fundus images, of which $50\%$ 
are used for training and $50\%$ are used for testing. Under the test data of STARE, an 
experimental result of CSAU is as follows: an F1-score of 0.8435, a SE of 0.8465 and an ACC of 0.9673.

In~\cite{Jiang-2020-multi}, a novel multi-path recurrent U-Net with attention gate (MPAR) is 
developed to segment retinal fundus images. An innovative idea of many variants of U-Net is 
that attention is integrated into the baseline 
U-Net~\cite{Lian-2018-AGU,Lv-2020-AGU,Mou-2019-CCA,Li-1903-CSA,Zhang-2020-PCS,Zhu-2020-SWR}). 
However, in MPAR, Attention Recurrent Unit (formed by combining recurrent neural network and 
attention) can further improve target features. Two datasets (Drishti-GS1 
dataset~\cite{Sivaswamy-2015-ACR} and DRIVE dataset~\cite{Bansal-2013-RVD}) are used to 
evaluate MPAR. The Drishti-GS1 dataset has 101 retinal fundus vascular images. For the 
Drishti-GS1 dataset, in the training phase, 50 images are used as the training set. In the 
testing phase, 51 images are used as the test set. Experimental results of testing MPAR under 
the DRISHTI-GS1 dataset are as follows. For optic disc segmentation, an ACC of $99.67\%$ and 
a DICE of $98.17\%$ are obtained. For optic cup segmentation, an ACC of $99.5\%$ and a DICE 
of $89.21\%$ are obtained.

\begin{figure}[htbp!]  
\centerline{\includegraphics[width=0.75\textwidth]{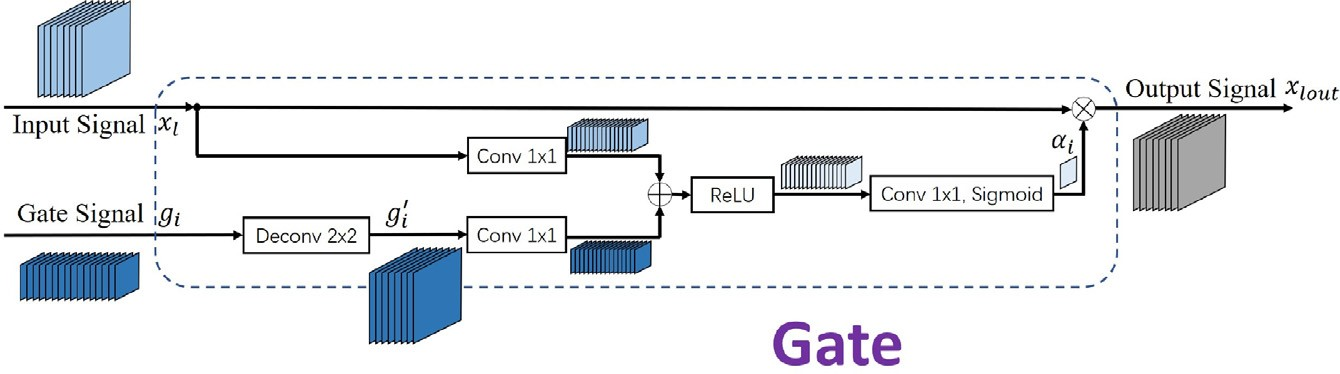}}
\caption{The network architecture of Attention Gate in~\cite{Zhang-2020-PCS} (Fig. 3).}
\label{Fig. 31}
\end{figure}

In~\cite{Zhang-2020-PCS}, a method based on ATT-UNet and graph-based Random Walk (RW) is 
proposed to extract nucleus and cytoplasm from overlapping cervical cells. This method proposed 
is mainly the following four steps: (1) ATT-UNet is used to separate the nuclei; (2) images are 
acquired by polar coordinate sampling; (3) ATT-UNet predicts the cytoplasm boundary; (4) RW is 
used to refine the cytoplasm boundary. Because of the repeated operation of the encoding part, 
some spatial detail information is lost. Attention Gates (AG)~\cite{Oktay-2018-AUL} is used 
to obtain the missing information. The architecture of  Attention Gate mentioned in 
Fig.~\ref{Fig. 31}. The experimental data are training images of the ISBI 2014 Challenge Dataset. 
135 synthetic cervical cytology images from eight training depth of field (EDF) images are 
splited into 45 images (train set) and 90 images (test set). Under the test set, a 0.93 DICE is 
obtained.

In~\cite{Zhu-2020-SWR}, irregular cell boundaries are often difficult to segment. Residual 
Attention U-Net (ResAttU-Net) is proposed to segment the cells in fluorescence widefield 
microscopy images. Like~\cite{Zhang-2020-PCS}, AG is applied to ResAttU-Net. 
Unlike~\cite{Zhang-2020-PCS}, Residual blocks~\cite{He-2016-DRL} are integrated into ResAttU-Net. 
This work utilizes a baby hamster kidney (BHK) cell dataset from Columbia University in the 
Department of Biomedical Engineering. After data enhancement such as random horizontal and 
vertical flipping, 4600 cell images constitute the training set, 1000 images constitute the 
training set and 500 images constitute the test set. This proposed method obtains good 
experimental results (SE=0.97, SP=0.93, F1-score=0.95, JS=0.91 and DICE=0.95).

In~\cite{Xiancheng-2018-RBV}, in order to assist ophthalmologists in the diagnosis of eye 
diseases, a method that relies on data enhancement and combined with U-Net is proposed to 
segment retinal blood vessel images. Like~\cite{Fabijanska-2018-SOC}, an identical structural 
deformation is applied to the baseline U-Net. Unlike~\cite{Fabijanska-2018-SOC}, this method 
is to segment the retinal blood vessels images. 20 DRIVE training images generate 19,000 patches 
to form a dataset. From the training set, $90\%$ of samples are used for training and the 
remaining $10\%$ are used for validation. This experiment obtains satisfactory results.

In~\cite{Leng-2018-CUF}, in order to fully consider contextual information, a context-aware 
U-Net is proposed to conduct the segmentation tasks of the Drosophila first instar larva VNC. 
Like~\cite{Chidester-2019-ERU}, the context-aware U-Net improves the long connection by placing 
a model on the connection. Unlike~\cite{Chidester-2019-ERU}, the context-aware U-Net places a 
context transfer module instead of a Residual module. A dataset used for the Drosophila first 
instar larva ventral nerve cord images segmentation in experiments 
from~\href{http://brainiac2.mit.edu/isbi_challenge/}{EM segmentation challenge (ISBI 2012)}. 
A training data is 30 serial section transmission electron microscopy images. The context-aware 
U-Net obtains a segmentation result: A warping error of 0.000121, a rand error of 0.0212 and 
a pixel error of 0.0346.

\subsection{Dense U-Net}
\begin{figure}[htbp!]  
\centerline{\includegraphics[width=0.65\textwidth]{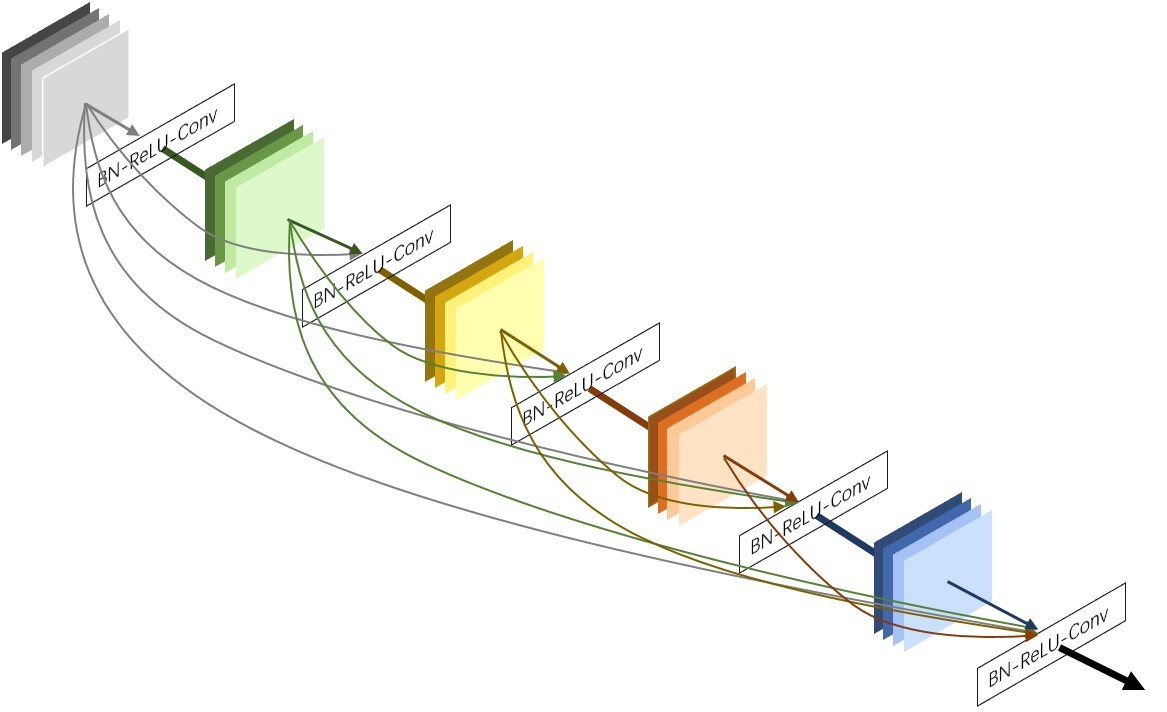}}
\caption{The network architecture of dense block in~\cite{Cheng-2020-RBV} (Fig.6). }
\label{Fig. 16}
\end{figure}

A densely connected convolutional network (DenseNet) is 
proposed~\cite{Huang-2017-DCC，Huang-2017-MDC}. The feature of a dense U-Net is that in the 
dense block, each layer is directly connected to the previous layer. The structure of the dense 
block is shown in Fig.~\ref{Fig. 16}.  The advantage of this structure is that the vanishing 
gradient problem can be alleviated. In~\cite{Jegou-2017-TOH}, DenseNet is modified and integrated 
into U-Net to propose a new neural network and used to segment some trial scenes. 

\begin{figure}[htbp!]  
\centerline{\includegraphics[width=0.75\textwidth]{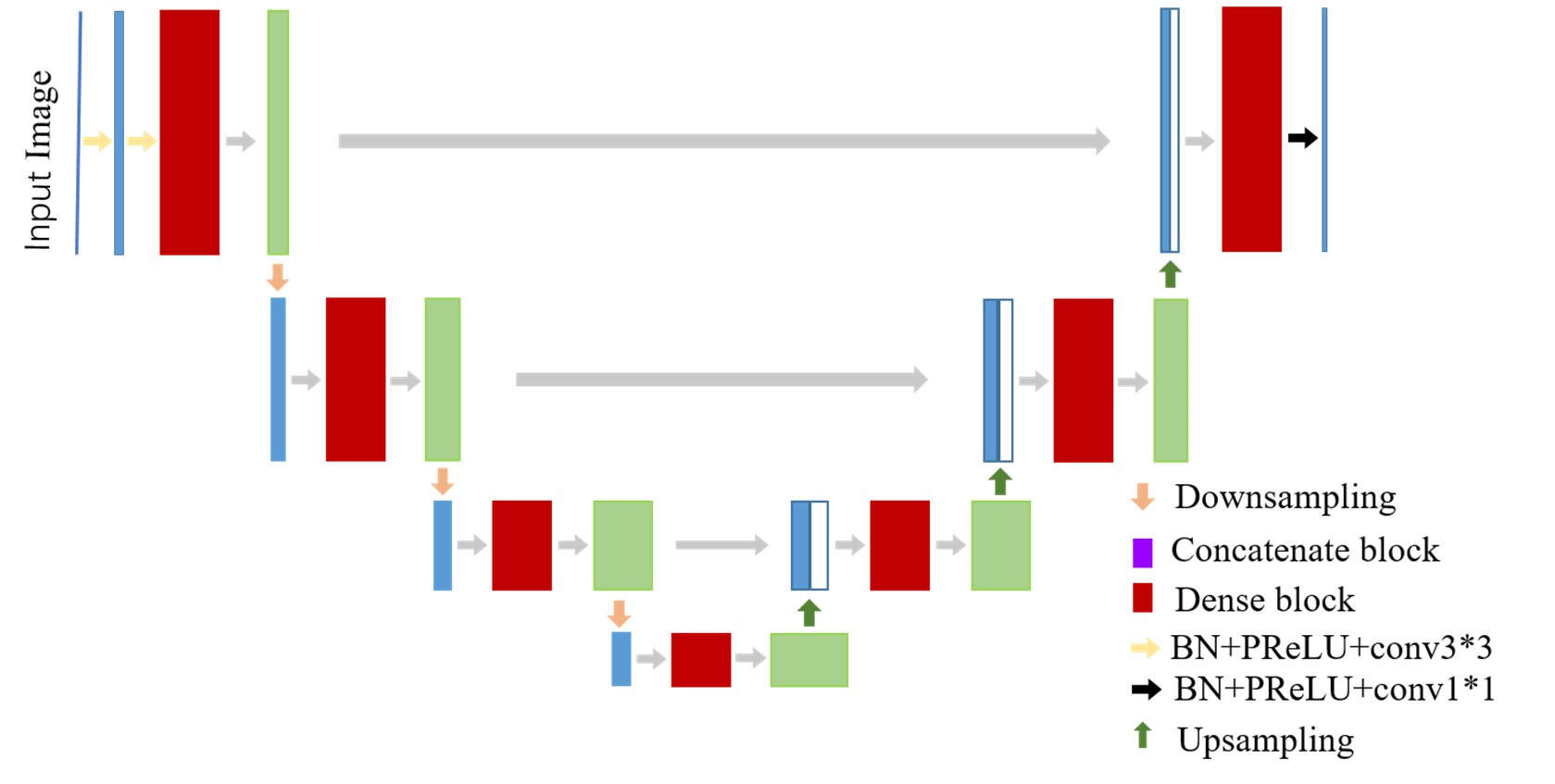}}
\caption{The network architecture of dense U-Net in~\cite{Cheng-2020-RBV} (Fig.7).}
\label{Fig. 17}
\end{figure}

In~\cite{Cheng-2020-RBV}, a new dense block-based U-Net is proposed to perform segmentation tasks 
of blood vessels. The structure of the new block-based dense U-Net is shown in Fig.~\ref{Fig. 17}. 
It differs from the baseline U-Net in: (1) dense block replaces ordinary convolutional blocks 
(in~\cite{Zhao-2020-TUH}, a PDFA module improves densely connected block); (2) Parametric 
Rectified Linear Unit (PReLU)~\cite{He-2015-DDI} replaces ReLU; (3) encoder and decoder part is 
missing a convolution block. two datasets (CHASE$\_$DB1 dataset~\cite{Owen-2009-MRV} and DRIVE 
dataset~\cite{Bansal-2013-RVD}) are used to evaluate the new dense block-based U-Net. 
The DRIVE dataset has 40 retinal fundus vascular images. For the DRIVE dataset from a diabetic 
fundus lesion screening organization, in the training phase, 20 images are used as the training set. 
In the testing phase, 20 images are used as the test set. Experimental results of testing the new  
block-based dense U-Net under the DRIVE dataset are as follows: 0.9834 SP, 0.7672 SE, 0.9559 ACC 
and 0.9793 AUC.

In~\cite{Wang-2019-DUB}, an improved U-Net combining patch-based learning strategy and dense 
idea is proposed to segment retinal vessels. Like~\cite{Cheng-2020-RBV}, compared with the 
baseline U-Net, dense block replaces ordinary convolutional blocks and the number of 
convolutional blocks is reduced. Unlike~\cite{Cheng-2020-RBV}, the Residual idea is applied 
before the dense block and the convolution block is reduced by two blocks in the encoder and 
decoder parts. This work utilizes two public datasets (DRIVE and STARE). For the DRIVE dataset, 
during the training process, 40000 image patches are extracted from 20 source images, of which 
$10\%$ are used for cross-validation. Finally, 0.7986 SE, 0.9736 SP, 0.9511 ACC and 0.9740 AUC 
are obtained.

In order to use deep learning models to assist pathologists in achieving precise treatment, 
in~\cite{Samanta-2021-CAN}, a new improved U-Net (HistNet) is proposed for segmentation of 
colorectal histopathology. Like~\cite{Cheng-2020-RBV,Wang-2019-DUB}, dense blocks are applied 
to U-Net. Unlike \cite{Cheng-2020-RBV,Wang-2019-DUB}, dense blocks are improved (modified dense 
block uses dilated convolution). To evaluate the proposed model thoroughly, the DigestPath 2019 
dataset~\cite{Li-2019-SRC} and the Gland Segmentation (GlaS) dataset (in Histology Image Challenge 
held at MICCAI 2015)~\cite{Sirinukunwattana-2017-GSI} are applied. For DigestPath 2019 dataset, 
it contains 660 tissue images from 324 WSI, the tissue images are randomly divided into training, 
validation and test sets at a ratio of $70:15:15$. Finally, $92.36\%$ DICE and $86.65\%$ IoU are 
obtained by HistNet.

Since, biomedical image segmentation plays an important role in diagnosing diseases, therefore, 
a novel Multi-scale Dense U-Net (MDU-Net) is proposed to segment biomedical 
images~\cite{Zhang-2018-MMD}. Like \cite{Cheng-2020-RBV,Wang-2019-DUB}, MDU-Net uses a dense idea.
Unlike~\cite{Cheng-2020-RBV,Wang-2019-DUB}, MDU-Net is a multi-scale densely idea, more precisely, 
Cross Dense connections, Up Dense connections, Down Dense connections are widely used in the 
MDU-Net, but MDU-Net does not have dense blocks. A dataset containing 165 biomedical images 
(originate from Histology Image Challenge held at MICCAI 2015) is used. The dataset is divided 
into two subsets, the first subset (containing 85 images) is used for training and the second 
subset (containing 85 images) is used for testing. Finally, experiment reveals that a DICE 
obtained by this proposed method is $4.1\%$ higher than baseline U-Net.

In~\cite{Liu-2018-DCS}, a Densely Connected Stacked U-Network (DCSU) is used to segment confocal 
microscopy images of filament. Unlike~\cite{Cheng-2020-RBV,Wang-2019-DUB,Zhang-2018-MMD}, DCSU is 
a cascaded U-Net (combination of multiple U-Nets, the output of previous level U-Net is related 
to the input of the next U-Net~\cite{Wu-2019-ABS}) and dense connections occur between convolutional 
blocks of different U-Nets. A microtubule dataset containing 5032407 training patches is proposed 
to evaluate DCSU. Under the microtubule dataset, this proposed method obtains an IoU of 0.9439 and 
a Skeletonized IoU (SKIoU) of 0.9775.

\subsection{ U-Net redesigned Skip Connections }
\begin{figure}[htbp!]  
 \centerline{\includegraphics[width=0.75\textwidth]{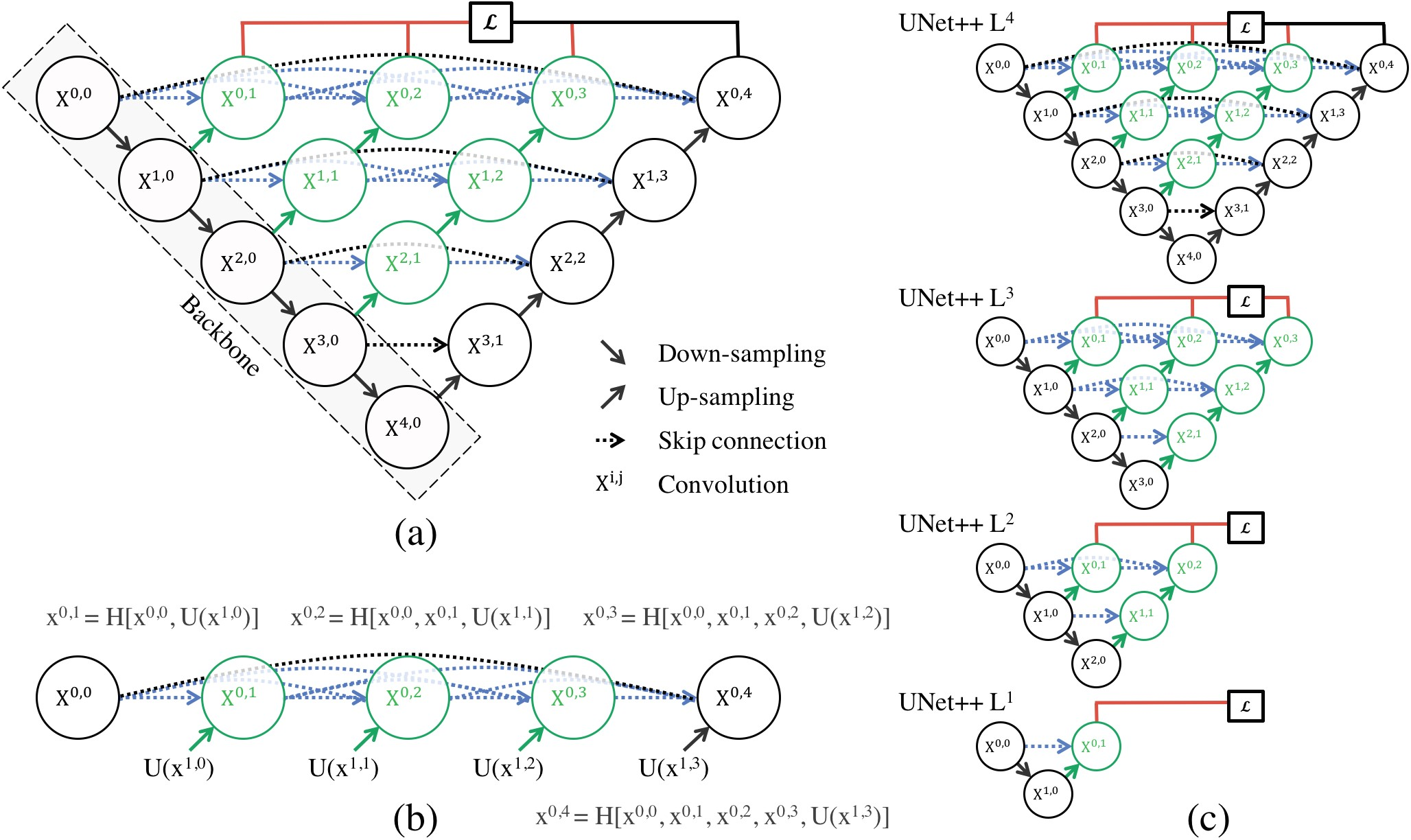}}
\caption{The network structure of the deformable U-Net in~\cite{Zhou-2018-UAN} (Fig.1).}
\label{Fig.8}
\end{figure}

In~\cite{Zhou-2018-UAN}, a simple and effective modified U-Net that redesigns the original 
U-Net skip connection is creatively proposed. It reduces the loss of information in the copy 
aggregation from the encoder to the decoder. For different datasets, the importance of each convolutional layer is different. The information carried by the third layer may be effective for 
the segmentation of the dataset or it may be the second layer. The advantage of this modified U-Net 
is that all copy aggregation operations contain all the depth feature information of the previous convolutional layer. A data from~\href{https://www.kaggle.com/c/data-science-bowl-2018}{Data 
Science Bowl 2018} containing 670 nuclear images are used. Under the dataset, an IoU of $92.63\%$ 
core segmentation result is obtained. Fig.~\ref{Fig.8} shows the proposed deformable U-Net and 
the analysis of the variant thinking. 

In~\cite{Zhou-2019-URS}, a U-Net that changes skip connection is explored in more depth 
(similarly, \cite{Zhou-2018-UAN} is the study of skip connection from the same team).
A purpose of the U-Net that changes the skip connection is to reduce the loss of information 
when convolutional layers of different depths are aggregated and connected. 
Like~\cite{Zhou-2018-UAN}, the deformed U-Net is deleted some mandatory aggregations (or 
different depth features are extracted), different performances are obtained. 
Unlike~\cite{Zhou-2018-UAN}, the deformed U-Net is evaluated by more datasets.

In~\cite{Wang-2020-AIB}, a powerful improvement U-Net++ is proposed to segment tiny breast 
cancer nuclei. The difference between the improved U-Net++ and the baseline 
U-Net++~\cite{Zhou-2018-UAN} is that an Inception-Resnet-V2 network is integrated into U-Net++ 
and the network improves U-Net++ segmentation capabilities. This work utilizes a dataset 
from~\cite{Janowczyk-2016-DLF}, which contains 141 RGB H$\&$E stained estrogen receptor 
positive (ER+) breast cancer images in total. Finally, the dataset has a total of 3366 
sub-images after being resized and cropped. The proportions of the training set, validation set 
and test set are $80\%$, $10\%$ and $10\%$, respectively. The improved U-Net++ is evaluated by 
this dataset and obtains 0.9505 ACC, 0.5581 PR, 0.6035 RE and 0.5207 DICE.

Doppler optical coherence tomography (OCT) vessels images can  observe the vascular structure 
and blood flow to facilitate surgery. So in~\cite{Wu-2019-ABS}, a cascaded U-Net (CU-net) is 
proposed to segment the vascular intensity images, the boundary image of the outer vessel wall 
and the inner blood flow lumen (this cascaded U-Net idea is also reflected in~\cite{Liu-2018-DCS}). 
In Cu-Net, the first U-Net segment the intensity image, then the segmentation result is used 
as the input of the second U-Net (as a mask, it is better to select the region of interest and 
remove the non-target region). The above operations can reduce training time. This method is 
examined on a traceable dataset of Doppler OCT images of mouse arteries. The dataset used in 
the experiments is consists of 190 images, of which 150 images are used for training and 40 
images are used for testing. These experimental results show that for the segmentation of the 
outer vessel wall boundary, CU-net obtains an ACC of $96.7\% \pm 0.2\%$, furthermore, for the segmentation of the contour of the inner blood flowing lumen area, CU-net obtains an ACC of 
$94.8\% \pm 0.2\%$.

In~\cite{Tsunomura-2020-SOM}, a dyed particle usually needs to be divided manually by an 
expert and the reproducibility of the operation is low. Therefore, a simple and effective 
deformed U-Net is proposed to solve the problem and achieve precise segmentation. A dataset is 
composed of art papers printed by mixing carbon black pigments and inks, with a total of 60 
images. First, improved U-Net $\#1$ is obtained by changing the number of channels. Then, 
improved U-Net $\#1$ outputs high-precision large particle segmentation results. Finally, 
the segmentation results are applied to improved U-Net $\#2$ (delete some skip connections) to 
obtain the best segmentation results. Two different variants of U-Net combined with the 
proposed method flow obtain a root mean squares error (RMSE) of 45.2187 and a standard 
deviation of 7.8371.

\begin{figure}[htbp!]  
 \centerline{\includegraphics[width=0.65\textwidth]{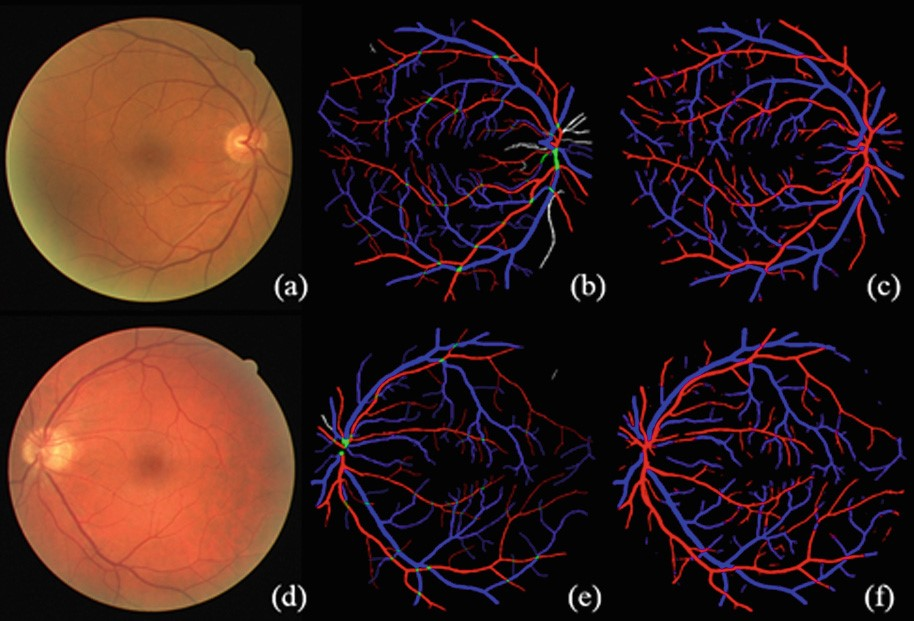}}
\caption{Segmentation result in~\cite{Xu-2018-AIU} (Fig.3). 
(a) and (d) are the original images; (b) and (e) are the ground truth images; 
(c) and (f) are the segmentation results.}
\label{Fig. 15}
\end{figure}

Hence, the segmentation of retinal arterioles and venules has an important auxiliary role 
in the diagnosis of eye diseases. Therefore, in~\cite{Xu-2018-AIU}, an improved U-Net is 
proposed to segment arterioles and venules. Like improved U-Net $\#2$ 
from~\cite{Tsunomura-2020-SOM} and improved U-Net from~\cite{Xu-2018-AIU} deletes the connections. 
On the contrary, improved U-Net $\#2$ from~\cite{Tsunomura-2020-SOM} and improved U-Net 
from~\cite{Xu-2018-AIU} only deletes the fourth connection between the encoder 512-channel 
feature map. The segmentation results of the proposed method are shown in Fig.~\ref{Fig. 15}. 
This experiment uses a DRIVE dataset curated by a diabetic retinopathy screening program. 
DRIVE (contains 40 color fundus photographs) is used to evaluate six different segmentation 
methods. Finally, the improved U-Net obtains an optimal experimental result (0.870 SE and 0.980 SP).

In~\cite{Liang-2020-WPM}, in order to better realize target spore identification and spore count, 
an effective variant U-Net is proposed to segment wheat powdery mildew spore images. The variant 
U-Net has two differences from the baseline U-Net as follows: The variant U-Net deletes the three 
skip connections in the baseline U-Net and adds the pyramid pool module after the encoding is 
completed. Like improved U-Net $\#2$ from~\cite{Tsunomura-2020-SOM} and improved U-Net 
from~\cite{Xu-2018-AIU}, the variant U-Net deletes some connections. Unlike improved U-Net $\#2$ 
from~\cite{Tsunomura-2020-SOM} and improved U-Net from~\cite{Xu-2018-AIU}, the variant U-Net 
from~\cite{Liang-2020-WPM} only 1 connection between encoder 512-channel feature map is reserved. 
835 wheat powdery mildew spore images are divided into a training set and a test set, of which 
550 images are used as training set and 285 images are used as the test set. The variant U-Net 
obtains an MIoU of $91.48\%$ experimental results.

\subsection{Summary}
From the survey above, since 2016, U-Net has improved, but after 2018, deep neural networks 
based on U-Net become more and more widely used in the field of biomedical image segmentation. 
The main reasons for this development trend are as follows: (a) More advanced computers are 
developed, which can handle more complex networks. (b) The improved U-Net architecture that 
can avoid problems such as over-fitting and enhanced computing time. Table~\ref{Table. 2} 
summarizes the work done by different teams in using the improved U-Net to analyze microscopic 
images.

\newcommand{\tabincell}[2]
{\begin{tabular}{@{}#1@{}}#2\end{tabular}}
\begin{table}[H]
\caption{Summary of the improved U-Net for segmentation tasks. 
The second column ``Detail'' shows the application object.}
\label{Table. 2}
\small 
\renewcommand\arraystretch{1.5}
\setlength{\tabcolsep}{1pt}          
\resizebox{\textwidth}{!}{
\begin{tabular}{|c|c|c|c|c|c|c|c|}
\hline
Aim                                                & Detial                      & Year & Reference                                    & Team                              & Data Information                                                                                                                                                 &   CNN type: points of improvement  & Evaluation                                                                                                                                                                                                                           \\ \hline
semi-automatic segmentation                        & Xenopus kidney slices       & 2016 & \cite{Cciccek-2016-3UN}     & A. Abdulkadir, et al.             & 77  images                                                                                                                                                       & 3D U-Net                                                                                                                                                                                                       & IoU = 86.3\%                                                                                                                                                                                                                         \\ \hline
segmentation                                       & 3D synthetic data           & 2018 &\cite{Fu-2018-TDF}          & C. Fu, et al.                     & 3D synthetic data                                                                                                                                                & 3D U-Net                                                                                                                                                                                                       & ACC = 95.56\%                                                                                                                                                                                                                        \\ \hline
segmentation                                       & cell membrane               & 2019 & \cite{Eschweiler-2019-CBP}  & D. Eschweiler, et al.             & \tabincell{c}{109296 cells for training,\\      972 cells for validation}                                                               & 3D U-Net combined with SWS                                                                                                                                                                                     & JI = 87\%, DICE = 93.1\%                                                                                                                                                                                                             \\ \hline
segmentation                                       & nerve tissue                & 2018 & \cite{Heinrich-2018-SCS}    & L. Heinrich, et al.               & \tabincell{c}{6 volumes: \\      75\% for  training, \\      25\% for validation}                                                       & 3D U-Net                                                                                                                                                                                                       & CREMI score = 50                                                                                                                                                                                                                     \\ \hline
segmentation                                       & neurons                     & 2019 & \cite{Wang-2019-SNS}        & H. Wang, et al.                   & \tabincell{c}{42 images: \\      38 images for training, \\      4 images for test}                                                  & \tabincell{c}{Teacher-student network: \\      based on 3D U-Net}                                                                                                                     & \textbackslash{}                                                                                                                                                                                                                     \\ \hline
\multirow{2}{*}{segmentation and   classification} & \multirow{2}{*}{SCD  cells} & 2017 & \cite{Zhang-2017-ISA}       & \multirow{2}{*}{M. Zhang, et al.} & 128 images                                                                                                                                                       & \multirow{2}{*}{variable convolution is applied}                                                                                                                                                               & \tabincell{c}{segmentation ACC = 97.8\%, \\      classification   ACC = 82.7\%}                                                                                                                             \\ \cline{3-4} \cline{6-6} \cline{8-8} 
                                                   &                             & 2018 & \cite{Zhang-2018-RSS}       &                                   & 266 original images                                                                                                                                              &                                                                                                                                                                                                                & \tabincell{c}{segmentation ACC =99.12\%, \\      IoU =  44.15\%}                                                                                                                                            \\ \hline
segmentation                                       & cell nuclei                 & 2020 & \cite{Qin-2020-MFU}         & X. F. Qin, et al.                 & \tabincell{c}{670 images:\\      80\%  for validation, \\      20\%  for test}                                                          & Match Feature U-Net                                                                                                                                                                                            & IoU = 91.83\%                                                                                                                                                                                                                        \\ \hline
segmentation                                       & trophectoderm               & 2020 & \cite{Rad-2020-TSI}  58     & R. M. Rad, et al.                 & \tabincell{c}{827 images:\\      70\% for training, \\      \%30 for test}                                                              & \tabincell{c}{dilated convolution;\\       Inceptioned modules}                                                                                                                       & \textbackslash{}                                                                                                                                                                                                                     \\ \hline
segmentation                                       & Rift V alley virus          & 2018 & \cite{Matuszewski-2018-MAT} & D. J. Matuszewski, et al.         & 143 TEM images                                                                                                                                                   & \tabincell{c}{reduction in the number of   feature\\      maps of the convolutional layer}                                                                                            & \tabincell{c}{DICE = 90\%, \\      IoU   = 83.1\%}                                                                                                                                                          \\ \hline
segmentation                                       & circulating tumor cells     & 2018 & \cite{Mocan-2018-ADO}       & I. Mocan, et al.                  & \tabincell{c}{56000 images:\\      70\% for training,\\       \%30 for test}                                                            & Convolution block increased                                                                                                                                                                                    & ACC = 99.81\%                                                                                                                                                                                                                        \\ \hline
segmentation                                       & endothelial cells           & 2018 & \cite{Fabijanska-2018-SOC}  & A. Fabijanska, et al.             & \tabincell{c}{30 images:\\      50\% for training, \\      50\% for test}                                                               & Convolution blocks are reduced                                                                                                                                                                                 & DICE = 86\%                                                                                                                                                                                                                          \\ \hline
segmentation                                       & nuclei                      & 2019 & \cite{Xu-2019-UFR}          & Z. Y. Xu, et al.                  & \tabincell{c}{878 patches:\\      650 patches  for training,\\        228 patches for test}                                             & \tabincell{c}{US-Net: \\      SSD  form a \\      post-processing sub-network}                                                                                                        & \textbackslash{}                                                                                                                                                                                                                     \\ \hline
automate segmentation                              & histopathological           & 2017 & \cite{Li-2017-NDL}          & W. Z. Li, et al.                  & \tabincell{c}{510 images:\\      110 images for training,\\       400 images for test}                                                  & \tabincell{c}{Convolution blocks are   reduced;\\      adds a noise-tolerant layer}                                                                                                   & \textbackslash{}                                                                                                                                                                                                                     \\ \hline
segmentation                                       & ventral nerve cord          & 2020 & \cite{kumar-2020-CSB}  67   & C. A. Kumar, et al.               & 120 images                                                                                                                                                       & \tabincell{c}{reducing the number of   filters;\\      adding a BN layer;\\      adding Adam}                                                                                         & IoU = 92.54\%                                                                                                                                                                                                                        \\ \hline
segmentation                                       & mitochondria and   synapses & 2018 & \cite{Bermudez-2018-ADT}    & R. Bermudez, et al.               & \textbackslash{}                                                                                                                                                 & 2  distributed U-Nets                                                                                                                                                                                        & JI = 72.3\%                                                                                                                                                                                                                          \\ \hline
segmentation                                       & retinal images              & 2018 & \cite{Zhuang-2018-LMN}      & J. T. Zhuang, et al.              & \tabincell{c}{28 retinal images:\\       70\% for training,  \\      30\% for test}                                                     & \tabincell{c}{LadderNet: two-stream\\       U-Net shares weights}                                                                                                                     & \tabincell{c}{SE = 79.78\%,\\      SP = 98.18\%,\\      ACC = 96.56\% \\      AUC = 98.39\%}                                                                                                                \\ \hline
segmentation                                       & Colorectal polyps           & 2020 & \cite{Jha-2020-DAD}         & D. Jha, et al.                    & CVC-ClinicDB dataset                                                                                                                                             & DoubleU-Net: the information flows from the   previous to the next decoding part                                                                                                                               & \tabincell{c}{RE = 84.57\%,\\      PR = 95.92\%}                                                                                                                                                 \\ \hline
segmentation                                       & adjacent cells              & 2020 & \cite{Torr-2020-DSO}        & A. Torr, et al.                   & \tabincell{c}{cells dataset : \\      50  images,\\      80\%  for training,  \\      10\%  for validation, \\        10\%    for test} & \tabincell{c}{DeepSplit:  one encoder branch \\      and two decoder branches}                                                                                                        & \tabincell{c}{DSC = 91.1\% \\      CDS = 90.3\%}                                                                                                                                                            \\ \hline
segmentation                                       & human skin                  & 2018 & \cite{Bozkurt-2018-AMC}     & A. Bozkurt, et al.                & \tabincell{c}{56 RCM mosaics: \\      46  for training, \\      10 for test}                                                            & Munet: nested with three   U-Nets                                                                                                                                                                              & \tabincell{c}{SE = 72.89\%,\\      SP = 95.26\%, \\      DICE = 78.36\%,\\      PR =84.71\%}                                                                                                         \\ \hline
segmentation                                       & nuclear                     & 2020 & \cite{Zhao-2020-TUH}  75    & B. Zhao, et al.                   & \tabincell{c}{30 images:\\      16 images  for training,\\       14 images  for test}                                                   & \tabincell{c}{three U-Net branches;\\      PDFA module is applied}                                                                                                                    & \tabincell{c}{AJI =  62.1\%,\\      DICE = 83.7\%,\\      Panoptic Quality = 60.1\%}                                                                                                                        \\ \hline
segmentation                                       & iris                        & 2018 & \cite{Lian-2018-AGU}        & S. Lian, et al.                   & \tabincell{c}{1000 images:\\      500 images for training,  \\      500 images for test}                                                & ATT-Unet: an attention mask is   generated                                                                                                                                                                     & IoU of 91.37\%                                                                                                                                                                                                                       \\ \hline
segmentation                                       & retinal blood vessels       & 2020 & \cite{Lv-2020-AGU}          & Y. Lv, et al.                     & \tabincell{c}{40 fundus images:\\      20 images  for training,\\       20 images  for test}                                            & \tabincell{c}{AA-Unet: attention guided   U-Net;\\      atrous convolution}                                                                                                           & \tabincell{c}{ACC = 95.58\%,  \\      F1-scores = 82.16\%, \\       JS = 95.68\%, \\      AUC = 98.47\%}                                                                                                    \\ \hline
segmentation                                       & retinal blood vessels       & 2019 & \cite{Mou-2019-CCA}         & L. Mou, et al.                    & STARE dataset                                                                                                                                                    & CS-Net: CAB and SAB is added                                                                                                                                                                                   & \tabincell{c}{ACC = 97.52\%,\\       SE = 88.16\%, \\      SP = 98.40\%, \\      AUC = 99.32\%}                                                                                                             \\ \hline
segmentation                                       & retinal blood vessels       & 2019 & \cite{Li-1903-CSA}          & R. Li, et al.                     & \tabincell{c}{20 fundus images:\\      50\%  for training, \\      50\%  for test}                                                      & CSAU: attention gates is added                                                                                                                                                                                 & \tabincell{c}{F1-score =84.35\%,  \\      SE = 84.65\%,  \\      ACC = 96.73\%}                                                                                                                             \\ \hline
segmentation                                       & retinal fundus              & 2020 & \cite{Jiang-2020-multi}     & Y. Jiang, et al.                  & \tabincell{c}{50  for training, \\      51  for test}                                                                                   & AG is added                                                                                                                                                                                                    & \tabincell{c}{For optic disc segmentation, an   ACC\\      of 99.67\% and a DICE of 0.9817 are obtained. For optic cup segmentation,   an\\     ACC of 99.5\% and a DICE of 0.8921 are obtained.} \\ \hline
segmentation                                       & cells                       & 2020 & \cite{Zhang-2020-PCS}       & H. Zhang, et al.                  & 135  images                                                                                                                                                      & AG is added                                                                                                                                                                                                    & \tabincell{c}{DICE\\      of 0.93}                                                                                                                                                                          \\ \hline
segmentation                                       & cell boundaries             & 2020 & \cite{Zhu-2020-SWR}         & N. Y. Zhu, et al.                 & \tabincell{c}{4600  images for training,\\       1000 images for validation,\\      500 images for test}                                & \tabincell{c}{AG is added;\\      Residual blocks is added}                                                                                                                           & \tabincell{c}{SE = 97\%,\\      SP = 93\%, \\      F1-score = 95\%,\\       JS = 91\%,\\      DICE=95\%}                                                                                                    \\ \hline
segmentation                                       & retinal blood   vessel      & 2018 & \cite{Xiancheng-2018-RBV}   & W. Xiancheng, et al.              & \tabincell{c}{19,000 patches,\\      90\% for training,  \\      10\%  for validation}                                                  & \tabincell{c}{An identical structural defor-\\      mation}                                                                                                                           & \textbackslash{}                                                                                                                                                                                                                     \\ \hline
segmentation                                       & l nerve                     & 2018 & \cite{Leng-2018-CUF}90      & J. X. Leng, et al.                & 30  images for  training                                                                                                                                         & \tabincell{c}{Context-aware U-Net: \\       placing a model on the connection}                                                                                                        & \tabincell{c}{a warping error = 0.000121,   \\      a rand error= 0.0212, \\      a pixel error = 0.0346}                                                                                                   \\ \hline
segmentation                                       & retinal fundus vascular     & 2020 & \cite{Cheng-2020-RBV}       & Y. L. Cheng, et al.               & \tabincell{c}{40 images,\\      20 images for training,  \\      20 images for test}                                                    & \tabincell{c}{dense block is added;\\      PReLU is added}                                                                                                                            & \tabincell{c}{SP = 98.34\%\\       SE = 76.72\%\\       ACC = 95.59\%   \\      AUC = 97.93}                                                                                                                \\ \hline
segmentation                                       & retinal vessels             & 2019 & \cite{Wang-2019-DUB}        & C. Wang, et al.                   & 40,000 image patches                                                                                                                                             & \tabincell{c}{dense block is applied;\\       convolutional blocks is   reduced;\\      Residual idea is applied}                                                                     & \tabincell{c}{SE = 79.86\%\\      SP =97.36\%\\       ACC = 95.11\%\\       AUC = 97.4\%}                                                                                                                   \\ \hline
segmentation                                       & colorectal histopathology   & 2021 & \cite{Samanta-2021-CAN}     & P. Samanta, et al.                & \tabincell{c}{660 tissue images,\\      70\% for training, \\      15\% for validation, \\      15\% for test}                          & \tabincell{c}{dense blocks is applied;\\       dilated convolution is applied}                                                                                                        & \tabincell{c}{DICE = 92.36\% \\      IoU = 86.65\%}                                                                                                                                                         \\ \hline
segmentation                                       & biomedical images           & 2018 & \cite{Zhang-2018-MMD}       & J. W. Zhang, et al.               & \tabincell{c}{165 biomedical images,\\      85 images for training,\\       85 images for test}                                         & \tabincell{c}{MDU-Net: a multi-scale densely   idea\\      Cross Dense connections is applied;\\       Up Dense connections is   applied;\\       Down Dense connections  is applied} & \textbackslash{}                                                                                                                                                                                                                     \\ \hline
segmentation                                       & filament                    & 2018 & \cite{Liu-2018-DCS}104      & Y. Liu, et al.                    & 5032407 training patches                                                                                                                                         & \tabincell{c}{DCSU: cascaded U-Net and \\      dense connections is applied}                                                                                                          & IoU = 97.75\%,                                                                                                                                                                                                                       \\ \hline
segmentation                                       & nuclear                     & 2018 & \cite{Zhou-2018-UAN}        & Z. W. Zhou,                       & 670  images                                                                                                                                                      & \tabincell{c}{redesigns the original \\      U-Net skip connection}                                                                                                                   & IoU of 92.63\%,                                                                                                                                                                                                                      \\ \hline
segmentation                                       & cell, nuclei                & 2019 & \cite{Zhou-2019-URS}        & Z. W. Zhou,                       &  & a new skip connection is applied                                                                                                                                                                               & \tabincell{c}{nuclei: IoU = 94\%,\\      DICE = 95.8\%}                                                                                                                                                     \\ \hline
segmentation                                       & nucle                       & 2020 & \cite{Wang-2020-AIB}        & H. Wang,                          & \tabincell{c}{3366 sub-images:\\      80\%  for training set, \\      10\% for validation,  \\       10\% for  test}                    & \tabincell{c}{Inception-Resnet-V2   network\\       is integrated\\      into U-Net++}                                                                                                & \tabincell{c}{ACC = 95.05\%, \\      PR = 55.81\%,\\      RE = 60.35\%, \\      DICE = 52.07\%}                                                                                                  \\ \hline
segmentation                                       & OCT vessels                 & 2019 & \cite{Wu-2019-ABS}          & C. C. Wu,                         & \tabincell{c}{190 images:\\       150 images  for training,\\       40 images  for test}                                                & \tabincell{c}{the first U-Net segment \\       intensity image,  the \\      segmentation result is used to \\      the second U-Net}                                                 & ACC = 94.8\% ± 0.2\%                                                                                                                                                                                                                 \\ \hline
segmentation                                       & pigments and inks           & 2020 & \cite{Tsunomura-2020-SOM}   & M. Tsunomura,                     & 60 images                                                                                                                                                        & \tabincell{c}{delete some skip connections;\\      changing the number of channels}                                                                                                   & \tabincell{c}{RMSE = 45.2187 \\       Standard deviation = 7.8371}                                                                                                                                          \\ \hline
segmentation                                       & color fundus photographs    & 2018 & \cite{Xu-2018-AIU}          & X. Y. Xu,                         & DRIVE: 40  images                                                                                                                                                & \tabincell{c}{deletes the fourth connection between \\      encoder’s 512-channel feature map}                                                                                        & \tabincell{c}{SE = 87\%\\       SP = 98\%}                                                                                                                                                                  \\ \hline
segmentation                                       & mildew spore                & 2020 & \cite{Liang-2020-WPM}114    & X. S. Liang                       & \tabincell{c}{835 images:\\      550 images for training, \\      285 images for test}                                                  & \tabincell{c}{deletes the three skip   connections;\\       pyramid pool module is applied}                                                                                           & MIoU = 91.477\%                                                                                                                                                                                                                      \\ \hline

\end{tabular}}
\end{table}

\newpage

\section{Modified U-Net Related to Residual Block and Residual Idea}
\label{Sec.4}
\subsection{Only Residual Block}
A depth of the network affects the extraction of features, but, as the network becomes deeper, 
the problem of overfitting and training errors in the network increases significantly. A Residual 
block is proposed to reuse the features of the previous layer to train a deeper 
network~\cite{He-2016-DRL}. In~\cite{Patel-2019-CSO}, an improved U-Net based on Residual blocks 
has introduced. It is used to segment the regions of human-induced pluripotent Retinal Pigment 
Epithelial stem cells (iRPE) under Bright-field microscopy. This improved U-Net is tested on 
1032 absorbance images of IRPE cells from Age-related Macular Degeneration (AMD) patients. In 
the dataset, the training set is composed of 800 images and the verification set is composed of 
232 images. The experiment shows that this method is superior to the prior art method and the 
proposed deformed U-Net obtains the experimental result of a DICE of 0.8366.

\begin{figure}[htbp!]  
 \centerline{\includegraphics[width=0.75\textwidth]{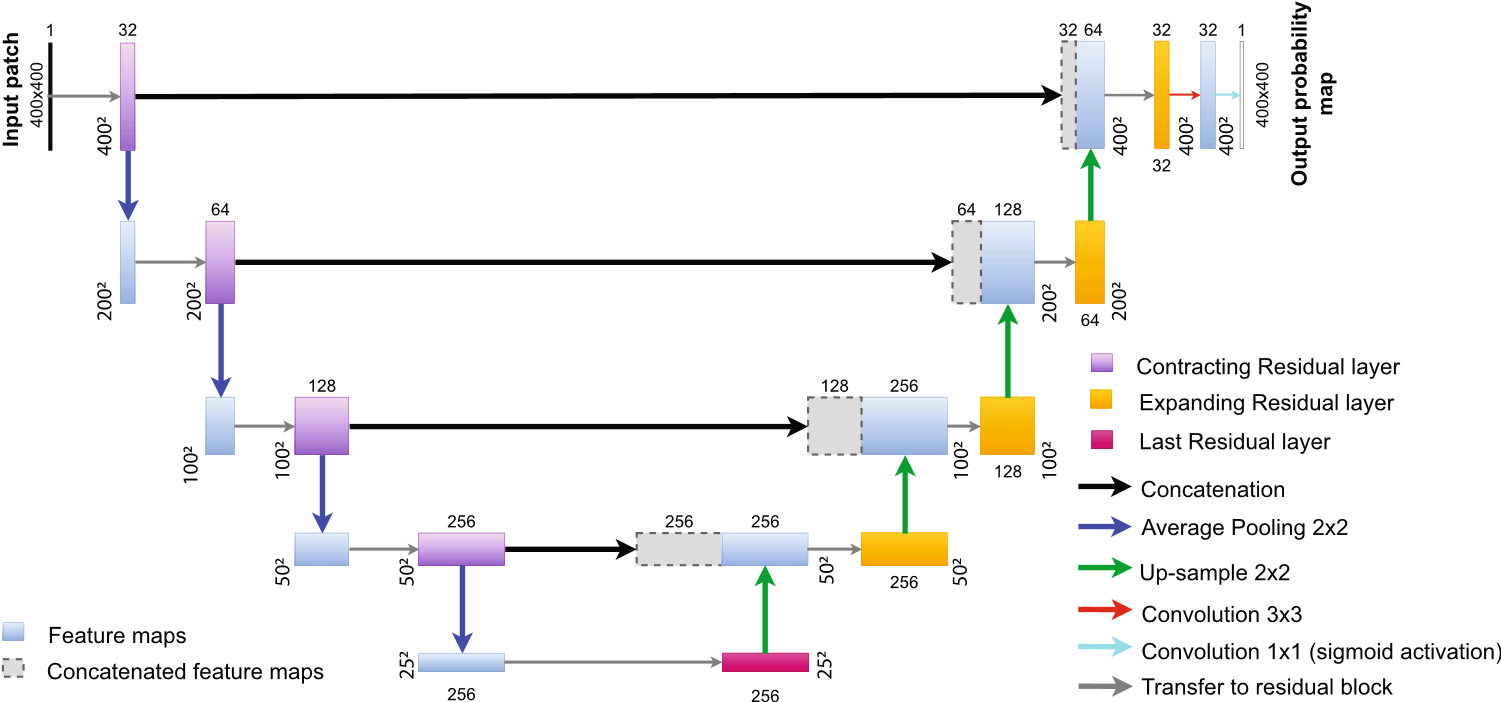}}
\caption{The network architecture of variant of U-Net in~\cite{Gomez-2019-DLB} (Fig.2).}
\label{Fig. 12}
\end{figure}

In~\cite{Gomez-2019-DLB}, a variant of U-Net with 9 Residual layers is proposed. It is used for 
Small extracellular vesicles (sEVs) collected by TEM to achieve good detection and segmentation 
performance. The architecture of variant  U-Net is shown in Fig.~\ref{Fig. 12}. 688 sEVs of Mouse 
fibroblasts (L-cells), mouse embryonic fibroblast (MEF), human embryonic kidney 293 (HEK-293) 
and the ovarian cancer cell are used as a dataset. Under the dataset, the proposed method obtains 
a JS of 0.88. 

In~\cite{Quan-2016-FAF}, FusionNet with Residual block based on U-Net is proposed to automatically 
segment neuronal cells. The biggest difference between FusionNet and the original U-Net is that 
a Residual block containing 3 convolutions is added to each block. 
Like~\cite{Patel-2019-CSO,Gomez-2019-DLB}, 3 variable U-Nets proposed in the three papers are 
all based on U-Net and add Residual blocks. 
Unlike~\cite{Patel-2019-CSO,Gomez-2019-DLB,Patel-2019-CSO,Gomez-2019-DLB} transformed the 
original convolution block into a Residual block, FusionNet retains the original convolution 
block and inserts Residual blocks with three convolutional. FusionNet is trained on 30 sections 
of Drosophila Electron Microscopic images (from~\cite{Arganda-2015-CTC}) and tested on 
a private dataset.

In \cite{Mehta-2018-YJS}, a simple concept network named Y-Net is used to segment different tissues 
in breast biopsy images to generate segmentation masks to assist breast cancer diagnosis. Y-Net 
differs from the baseline U-Net in that: Pyramid spatial pooling (PSP) and Residual convolutional 
blocks (RCB) are integrated into Y-Net, Y-Net adds two new hyperparameters and new fully connected 
layers.

The breast biopsy dataset containing 58 regions of interest (ROIs) is divided into a training 
set containing 29 images and a test set containing 29 images. Y-Net obtains a mIoU of $44.19\%$ 
experimental results.

However, a large number of manual annotations are time-consuming and labor-intensive, a 
Multi-Tasking U-Net is proposed to solve this problem~\cite{Ke-2019-AMU}. The Multi-Tasking 
U-Net is trained by coarse data labels to combine with a few pixel-wise annotations images. 
A Residual Multi-Tasking Block is proposed, each Multi-Tasking Block has three paths. The 
Multi-Tasking Block is composed of several sub-blocks with three tasks: task one is detection, 
task two is separation and task three is segmentation. Like~\cite{Torr-2020-DSO}, the 
Multi-Tasking U-Net is also a Multi-Tasking network. Unlike~\cite{Torr-2020-DSO}, the 
multi-tasking process of Multi-Tasking U-Net is embodied in a module instead of a decoder and 
Multi-Tasking U-Net has 1 more detection task. A training dataset contains 20 ice-cream Scanning 
Electron Microscopy (ESM) images and a test dataset contains 12 ice-cream SEM images. This 
Multi-Tasking U-Net obtains a DICE of 0.94 experimental results.

\subsection{Residual Block $\&$ Skip Connection}
\begin{figure}[htbp!]  
 \centerline{\includegraphics[width=0.75\textwidth]{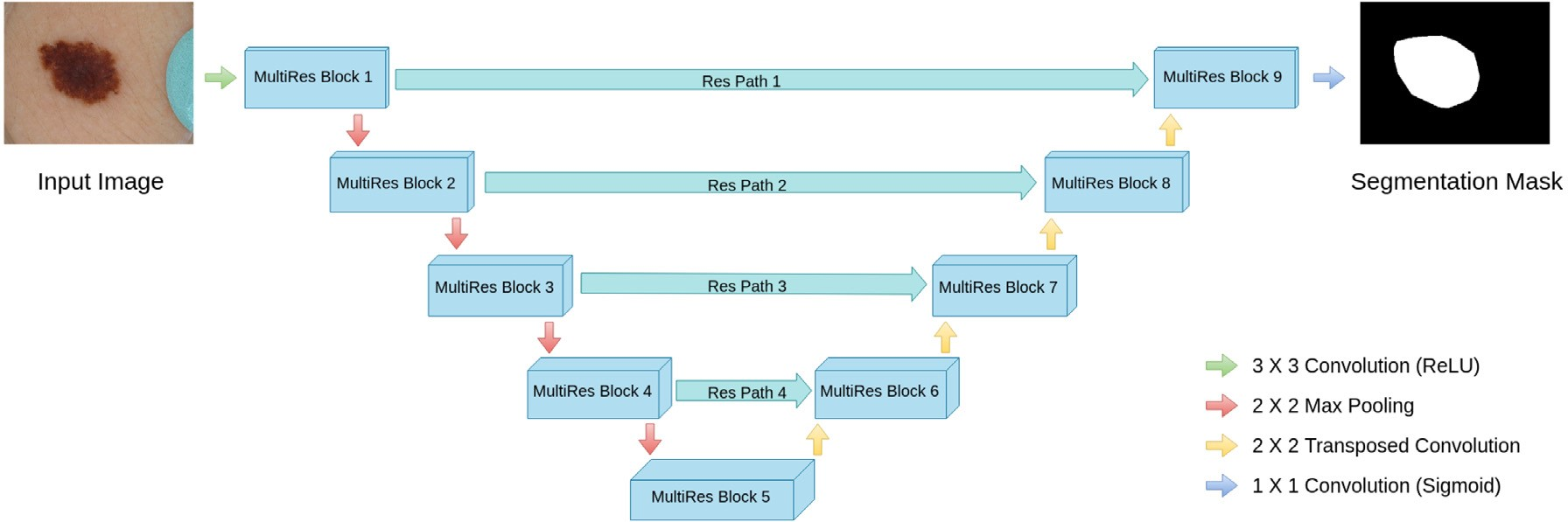}}
\caption{The network architecture of MultiResUNet in~\cite{Ibtehaz-2020-MRT} (Fig.5).}
\label{Fig. 13}
\end{figure}

In~\cite{Ibtehaz-2020-MRT}, in order to solve a problem that the classical U-Net does not 
perform well in the segmentation of some challenging datasets, MultiResUNet is proposed. 
Compared with the classical U-Net, the improved performance of MultiResUNet lies in the following 
two points. First, the MultiRes block is proposed and replaces all 2D convolution blocks in the 
classical U-Net. Second, a Res path with four $3 \times 3$ convolution and four short connections 
of Residual properties are creatively proposed to replace skip connections. MultiResUNet is tested 
and evaluated on five datasets, one of them obtains a JI of $91.65\%$  experimental results (the 
dataset containing 97 fluorescence microscopy images from~\cite{Coelho-2009-NSI}). This architecture 
of MultiResUNet mentioned is shown in Fig.~\ref{Fig. 13}.

In~\cite{Lou-2021-DRT}, to minimize the limitations of the classical U-Net in some aspects and 
obtain more accurate segmentation results, DC-UNet is proposed. Like~\cite{Ibtehaz-2020-MRT}, 
the network structure of both uses the same Res path instead of skip connection. 
Unlike~\cite{Ibtehaz-2020-MRT}, DC-UNet creates a Dual-Channel block (consisting of six 
$3 \times 3$ Residual convolutions in two rows) to replace all convolution blocks of the original 
U-Net. The dataset from the ISBI 2012 challenge (a training set of 30 images and a test set of 30 
images) is used to evaluate DC-UNet. An experiment is carried out 5-fold cross-validation and 
the experimental result of an average ACC of $92.62\%$ is obtained.

In~\cite{Gadosey-2020-SSD}, in order to segment biomedical images quickly, effectively and with 
high PR, SD-UNet is proposed. It has three advantages: small model size (23 times smaller than 
U-Net), fewer parameters (eight times less than U-Net) and fast computing time. Compared with 
U-Net, its change is that a SD-UNet block is designed and replaces the original convolution 
block. From International Symposium on Biomedical Imaging (ISBI) challenge dataset, 30 fruit fly 
images under ssTEM are used to test SD-UNet to obtain an IoU of $83.26\%$ and a DICE of $97.84\%$.

In~\cite{Arbelle-2019-MCS}, a Convolutional Long Short Term Memory (C-LSTM) block is integrated 
into U-Net and LSTM U-Net is proposed. The limitation of time information is incorporated into 
LSTM U-Net. Like~\cite{Gadosey-2020-SSD}, a new module is designed to replace the original 
convolution module. Unlike~\cite{Gadosey-2020-SSD}, the C-LSTM blocks of LSTM U-Net only exist 
in the encoder part. The structure facilitates the segmentation of single touch cells and 
partially visible cells. Furthermore, similar to an application of the LSTM module is discussed 
in this paper~\cite{Abdallah-2020-RAR}, which an extended LSTM block is designed based on the 
LSTM module and a Res block is designed based on the baseline U-Net and Residual. The LSTM U-Net 
is tested by the fluorescent simulated dataset (Fluo-N2DH-SIM+) 
in~\href{http://www.celltrackingchallenge.net.}{the Cell Tracking Challenge} and a quantitative 
result of 0.811 is obtained.

In~\cite{Chidester-2019-ERU}, since U-Net still has room for improvement in the task of nuclear 
segmentation, Rotation-Equivariant U-Net (REU-Net) is proposed to be used for histopathological 
images of seven different organs. Like~\cite{Ibtehaz-2020-MRT,Gadosey-2020-SSD,Lou-2021-DRT}, 
the REU-Net adds Residual blocks and modifies long connection. 
Unlike~\cite{Ibtehaz-2020-MRT,Gadosey-2020-SSD,Lou-2021-DRT}, in REU-Net, a long connection 
is between the second encoder and the third encoder. The dataset contains of 30 pathological 
images curated by~\cite{Kumar-2017-ADA}. In the dataset, 4 images are used as the training set 
and 7 images are used for validation and testing set respectively. The proposed method obtains 
an experimental result of an aggregated JS of 0.6291, an F1-score of 0.8469 and a DICE of 0.7980.

\subsection{Residual Block $\&$ Recurrent}
Models such as CNN can not accurately simulate high-level dependencies between object boundary 
points. Furthermore, in order to prevent overfitting and reduce the computational time, 
Recurrent Active Contour Evolution Network (RACE-Net) is proposed to segment Optic disc (OD) 
and Optic Cup (OC) in fundus images~\cite{Chakravarty-2018-RAR}. Feedforward neural network (FFNN) 
architecture is added to RACE-Net to simulate every step of the curve evolution. A generalized
the level set based deformable models (LDM) evolving is simulated by RACE-net. RACE-Net utilizes 
the DRISHTI-GS1 dataset from~\cite{Sivaswamy-2015-ACR}, which contains 101 images in total (50 
for training and 51 for testing). This RACE-Net-based method has the following experimental 
results. For OD segmentation, a DICE of 0.97 and a Boundary Localization Error (BLE) of 6.06 
are obtained. For OC segmentation, a DICE of 0.87 and a BLE of 7.63 are obtained.

\begin{figure}[htbp!]  
\centerline{\includegraphics[width=0.75\textwidth]{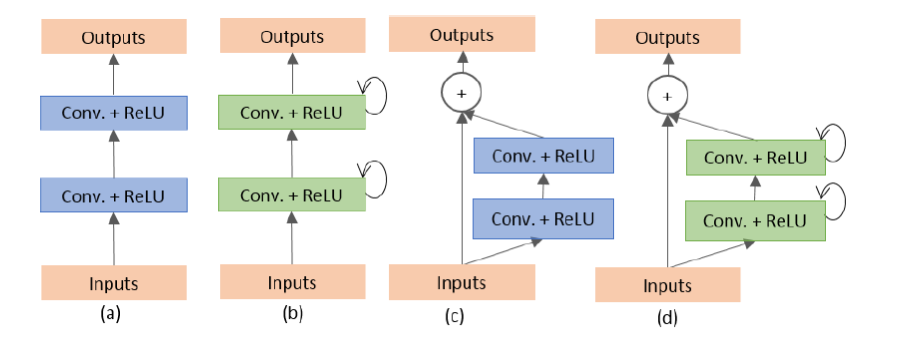}}
\caption{Different variant of convolutional and recurrent convolutional units 
in~\cite{Alom-2018-RRC} (Fig.4). (a) Forward convolutional units; 
(b) Recurrent convolutional block; (c) Residual convolutional unit; 
(d) Recurrent Residual convolutional units (RRCU). }
\label{Fig. 41}
\end{figure}

In~\cite{Alom-2018-RRC,Alom-2019-RRU}, Recurrent U-Net (RU-Net) and Recurrent Residual U-Net (R2U-Net)  
are developed to accurately segment the blood vessel images on the retina and skin cancer images. 
The biggest difference between RU-Net, R2U-Net and the baseline U-Net~\cite{Ronneberger-2015-UCN} 
is that the convolution units are different. RU-Net uses recurrent convolutional layers (RCL) and 
R2U-Net uses recurrent Residual convolutional layers (RRCL). Different variants of convolutional 
and recurrent convolutional units are shown in Fig.~\ref{Fig. 41}. The dataset contains 25000 
patches of 20 images from Structured Analysis of the Retina (STARE) and CHASE$\_$DB1 (of which 
$90\%$ is train set and $10\%$ is validation set) is used to train and verify RU-Net, R2U-Net.
RU-Net obtains experimental results of an F1-score of 0.8396, a SE of 0.8108, an SP of 0.9871, 
an ACC of 97.06\% and an AUC of 0.9909, respectively. R2U-Net obtains the experimental results 
of an F1-score of 0.8475, a SE of 0.8298, an SP of 0.9862, an ACC of 97.12\% and an  AUC of 
0.9914, respectively.

In~\cite{Alom-2018-NSW} (from the same team as~\cite{Alom-2018-RRC}), R2U-Net is applied to the 
segmentation task of a nuclear cell. An unfolded version of the recurrent convolutional units 
is proposed. Like~\cite{Alom-2018-RRC}, R2U-Net from~\cite{Alom-2018-NSW} has the same structure.  
Unlike~\cite{Alom-2018-RRC}, R2U-Net from~\cite{Alom-2018-NSW} is applied to nuclear segmentation 
for the first time. This R2U-Net is evaluated on the Data Science Bowl Grand Challenge in 2018 
(a total of 735 images, of which 536 are used for training, 134 are used for validation and 65 
are used for testing). Finally, an experimental result shows that a DICE of $92.15\%$ is obtained 
on the test set.

In~\cite{Zahangir-2018-MNC}, an R2U-Net model is developed to segment the nucleus. R2U-Net is 
introduced in~\cite{Alom-2018-RRC}, compared with R2U-Net from~\cite{Alom-2018-RRC}, this R2U-Net 
has not changed at all. A dataset used for image segmentation in experiments is composed of 735 
cell images from the 2018 Data Science Bowl Grand Challenge. The database is divided into two 
groups: training and testing. The training dataset includes 650 images (in the training set, 
$80\%$ is used for training and $20\%$ is used for validation), while the testing dataset 
includes 65 images. Finally, this nuclei segmentation task obtains $92.15\%$ testing ACC.

In~\cite{Yang-2020-ESA}, a baseline U-Net is introduced to segment label-free multiphoton 
microscopy (MPM) images of epithelial cells in prostate tissue. Segmentation results (segmented 
by the baseline U-Net) combined with the input of the baseline U-Net to obtain a merged image to 
train AlexNet~\cite{Krizhevsky-2012-ICW} for classification. Nine tissue slides and 70 tissue 
microarray (TMA) cores of prostate cancer (PCa) tissues from 79 patients are used as a dataset. 
The dataset is curated by the First Affiliated Hospital of Fujian Medical University. Segmentation 
results are obtained: A mean F1-score of 0.839.

\subsection{Inception-ResNet Block}
Inception~\cite{Szegedy-2017-IIA} and ResNet~\cite{He-2016-DRL} receive widespread attention 
since they are creatively proposed. Different receptive fields are the characteristics of 
Inception structures and ResNet has a unique connection method to avoid vanishing gradients. 
In~\cite{Szegedy-2017-IIA}, Inception and ResNet are combined to form an Inception-ResNet 
block that combines the advantages of the above two structures.

In order to solve the problem of spatial information loss caused by continuous pooling and 
convolution in U-Net, a context encoder network (CE-Net) is proposed to segment medical 
images~\cite{Gu-2019-CCE}. By combining Inception-ResNet-V2 block and atrous convolution, dense 
atrous convolution (DAC) blocks are proposed. The structural improvements are: (1) in the coding 
part, pre-trained ResNet-34~\cite{He-2016-DRL} replaces original ordinary blocks; (2) DAC blocks 
and Residual multi-kernel pooling (RMP) blocks are inserted into the context extractor section. 
This work utilizes an ORIGA dataset from~\cite{Zhang-2010-OAO}, which contains 650 optic disc 
images in total. The proportions of the training set and test set are $50\%$ and $50\%$, 
respectively. Under the dataset, CE-Net obtains an experimental result of an overlapping error 
of 0.058.

\begin{figure}[htbp!]  
\centerline{\includegraphics[width=0.5\textwidth]{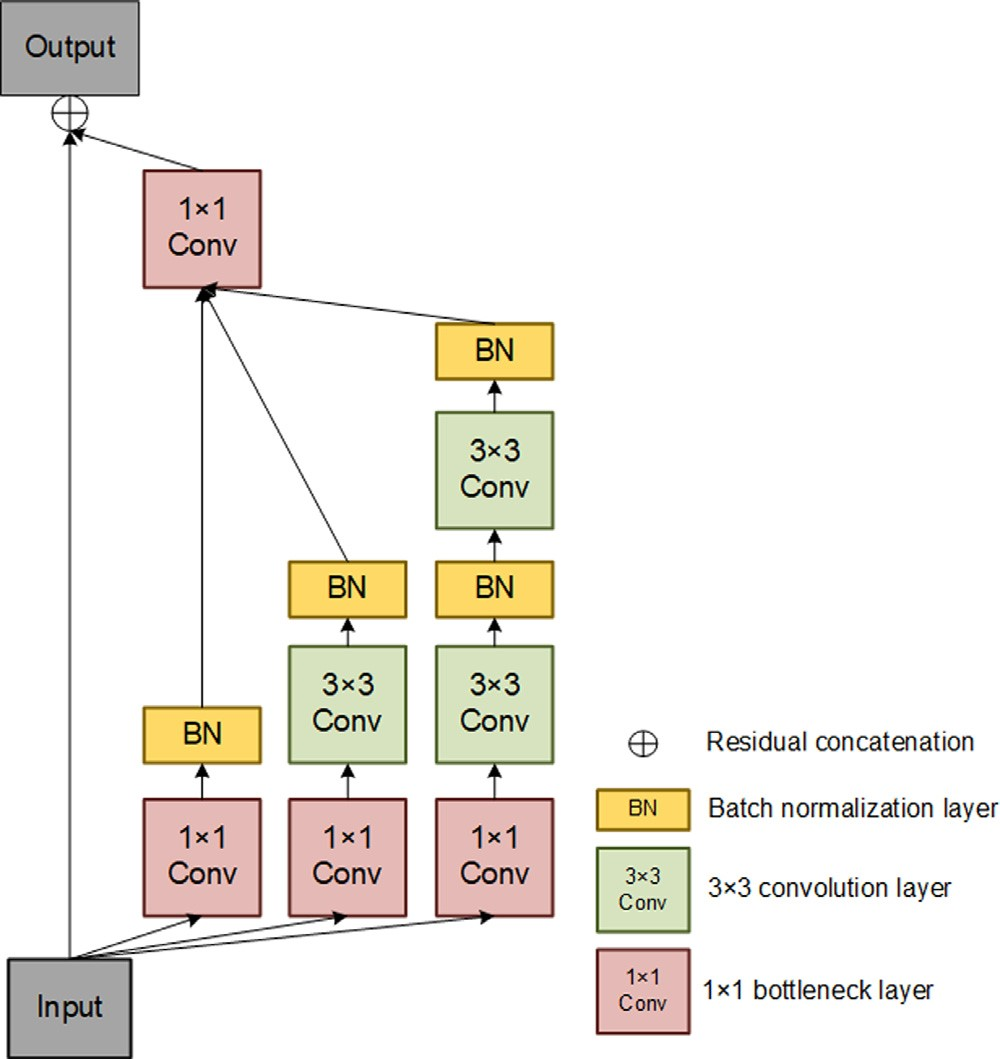}}
\caption{The network architecture of Inception-Res block in~\cite{Zhang-2020-DUF} (Fig.2).}
\label{Fig. 18}
\end{figure}

In order to solve the problem of vanishing gradient and excessive computation, 
in~\cite{Zhang-2020-DUF}, a Dense-Inception U-net (DIU-Net) with an Inception-res block is 
proposed to segment blood vessels. Like~\cite{Gu-2019-CCE}, Residual (short connection) and 
Inception are combined to form an Inception-Res block (The structure of the Inception-Res block 
is shown in Fig.~\ref{Fig. 18}). Unlike~\cite{Gu-2019-CCE}, the Dense-Inception block is applied 
in DIU-Net and the baseline Inception-Res block from~\cite{Szegedy-2017-IIA} is replaced by a 
modified Residual Inception module. Similar to~\cite{Lv-2020-AGU}, three datasets 
(DRIVE~\cite{Bansal-2013-RVD}, STARE~\cite{Guo-2018-ARV} and CHASH$\_$DB1~\cite{Thangaraj-2018-RVS}) 
are used to evaluate the improved U-Net.  The dataset has 136 samples of RGB images, of which 
$85\%$ is used for training and $15\%$ is used for testing. five-fold cross-validation is used 
in this experiment. Finally, a good experimental result is obtained (0.9582 DICE, 0.9338 JS, 
0.9657 ACC, 0.7967 SE, 0.9863 SP, F0.8003 1-score and 0.9802 AUC). 

In order to better distinguish tiny features between different categories, in~\cite{Huang-2020-MFM}, 
a Mini-Inception-Residual-Dense network (MIRD-Net) is proposed to segment cervical cancer cells, 
blood vessels, and nuclei. Like~\cite{Gu-2019-CCE,Zhang-2020-DUF}, MIRD-Net uses Inception blocks 
and Residual blocks. Unlike~\cite{Gu-2019-CCE,Zhang-2020-DUF}, MIRD-Net integrates the Inception 
block, Residual block and dense block into a Mini-Inception-Residual-Dense Block, instead of being 
used independently for the network. A dataset of 30 images of nuclei from Data Science bowl 2018 
is used. Under 5-fold cross-validation, an experimental result of a DICE of 0.954 is obtained.

In~\cite{Trimeche-2020-FAC}, an improved U-Net is proposed to automatically segment retinal vessel 
branches and bifurcations. In \cite{Trimeche-2020-FAC}, the baseline convolution block 
from~\cite{Ronneberger-2015-UCN} is replaced by Residual Inception blocks 
from~\cite{Szegedy-2017-IIA}. Different from~\cite{Gu-2019-CCE,Zhang-2020-DUF,Huang-2020-MFM}, 
Firesqueeze blocks are inserted into U-Net. The improved U-Net is evaluated on a private dataset 
containing 65 Adaptive Optics Ophthalmoscopy (AOO) images of the eye fundus. In the dataset, 
30 images are selected as the training set, 5 images are selected as the validation set and 
30 images are selected as the test set. Finally,  experiments reveal that the improved U-Net 
obtains a PR of 0.97, a RE of 0.96 and an F1-score of 0.96.

\subsection{Summary}
According to the review above, we can see that, since 2018, modified U-Nets related to Residual 
block and Residual ideas are widely used. The variant U-Net without the Residual idea has some 
limitations, such as easy over-fitting and slack training speed, while the Residual U-Net has 
better encode and decode information. The combination of Residual ideas and Inception, recurrent, 
skip connection changes can perform their respective advantages and the image segmentation 
efficiency is better. Table~\ref{Table. 3} summarizes the work done by different teams in using 
modified U-Net related to Residual block and Residual idea U-Net to analyze microscopic images.
 \begin{table}[htbp!]
\caption{Summary of the modified U-Net related to Residual block and Residual idea U-Net for segmentation tasks. The second column  ``Detail'' shows the application object.}
\label{Table. 3}
\small 

\renewcommand\arraystretch{1.5}
\setlength{\tabcolsep}{1pt}          
\resizebox{\textwidth}{40mm}{

\begin{tabular}{|c|c|c|c|c|c|c|c|}
\hline
Aim          & Detial                                   & Year & Reference                                                                      & Team            & Data Information                                                                                                                                                  & CNN type: points of improvement   & Evaluation                                                                                                                                                                                        \\ \hline
segmentation & Irpe stem cells                          & 2019 & \cite{Patel-2019-CSO}                                         & G. Patel, et al.                    & \tabincell{c}{800 images\\       232 for validation .}                                                                            & Residual blocks instead of ordinary blocks                                                                                                                                             & DICE = 83.66\%                                                                                                                                                                                    \\ \hline
segmentation & sEVs                                     & 2019 & \cite{Gomez-2019-DLB}                                         & E. Gomez de Mariscal, et al.        & 688  images                                                                                                                                                                              & 9 Residual layers instead of ordinary blocks                                                                                                                                        & JI = 88\%                                                                                                                                                                                         \\ \hline
segmentation & neuronal cells                           & 2016 & \cite{Quan-2016-FAF}                                          & T. M. Quan, et al.                  & 30    for  test                                                                                                                                                                   & \tabincell{c}{FusionNet: Residual block   containing 3 convolutions is\\      added to each block}                                                            & \tabincell{c}{Vrand = 97.8\%\\      Vinfo = 98.99\%}                                                                                                                     \\ \hline
segmentation & tissues in breast biopsy                 & 2018 & \cite{Mehta-2018-YJS}                                         & S. Mehta, et al.                    & \tabincell{c}{58 ROIs\\      29  for training\\   29   for test}                                                                                    & \tabincell{c}{Y-Net: adds PSP and RCB;\\       adds two new hyperparameters;\\       adds a new fully connected layers}                                       & mIoU = 44.19\%                                                                                                                                                                                    \\ \hline
segmentation & ice cream                                & 2019 & \cite{Ke-2019-AMU}120                                         & R. Ke, et al.                       & \tabincell{c}{20      training\\      12  for test}                                                                                                 & Multi-Task U-Net: each Multi-Task Block has three paths                                                                                                                                & DICE  = 0.94\%                                                                                                                                                                                    \\ \hline
segmentation & fluorescence microscopy   images         & 2020 & \cite{Ibtehaz-2020-MRT}                                       & N. Ibtehaz, et al.                  & 97 fluorescence microscopy images                                                                                                                                                        & \tabincell{c}{Residual block is added;\\       Skip connection is changed}                                                                                    & JI = 91.6537\%                                                                                                                                                                                    \\ \hline
segmentation & cells                                    & 2021 & \cite{Lou-2021-DRT}                                           & A. Lou, et al.                      & \tabincell{c}{ISBI-2012 challenge\\ 30  for training\\   30  for test                                                                                                                       } & \tabincell{c}{Residual block is added;\\       Skip connection is changed}                                                                                    & ACC = 92.62\%                                                                                                                                                                                \\ \hline
segmentation & neuronal cells                           & 2020 & \cite{Gadosey-2020-SSD}                                       & P. K. Gadosey, et al.               & \tabincell{c}{ISBI challenge\\30 fruit fly images                                                                                                                                                     } & \tabincell{c}{Residual block is added;\\       Skip connection is changed}                                                                                    & \tabincell{c}{IoU = 83.26\%\\      DICE = 97.84\%}                                                                                                                       \\ \hline
segmentation & fluorescence microscopy   images         & 2019 & \cite{Arbelle-2019-MCS}                                       & A. Arbelle, et al.                  & \tabincell{c}{fluorescent simulated dataset   \\Fluo-N2DH-SIM+                                                                                                                                     } & \tabincell{c}{Residual block is added;\\       Skip connection is changed}                                                                                    & Quantitative results = 81.1\%                                                                                                                                                                     \\ \hline
segmentation & nuclear                                  & 2019 & \cite{Chidester-2019-ERU}91                                   & B. Chidester, et al.                & \tabincell{c}{30 pathological images\\       4 for  training\\ 7  for validation\\ 7 images for test}                                   & \tabincell{c}{Residual block is added;\\       Skip connection is changed}                                                                                    & \tabincell{c}{AJI =62.91\%,\\      F1-score = 84.69\%,\\      DICE =79.80\%}                                                                                             \\ \hline
segmentation & OD and OC                                & 2018 & \cite{Chakravarty-2018-RAR}                                   & A. Chakravarty                      &\tabincell{c}{101 images\\      50 for training, 51 for test}                                                                                                   & \tabincell{c}{RACE-Net: Recurrent;\\      FFNN architecture is added;\\      LDM is added}                                                                    &\tabincell{c}{DICE = 0.87\%,\\      BLE = 7.63}                                                                                                                          \\ \hline
segmentation & blood vessel                             & 2018 & \cite{Alom-2018-RRC},   \cite{Alom-2019-RRU} & M. Z. Alom, et al. & \tabincell{c}{25000 patches \\      90\% for training, 10\% for validation}                                                                                     & RU-Net and R2U-Net (use RCL and   RRCL)                                                                                                                                                & \tabincell{c}{F1-score = 84.75\%, \\      SE = 82.98\%,  \\      SP = 98.62\%, \\      ACC = 97.12\%,\\      AUC = 99.14\%}                                              \\ \cline{1-4} \cline{6-8}
segmentation & nuclear cell                             & 2018 & \cite{Alom-2018-NSW}                                          &                                     & \tabincell{c}{735 images\\       536 for training, 134 for validation   and 65  for test}                                                                       & R2U-Net                                                                                                                                                                                & DICE = 92.15\%                                                                                                                                                                                    \\ \hline
segmentation & nucleus                                  & 2018 & \cite{Zahangir-2018-MNC}                                      & A. M. Zahangir, et al.              & \tabincell{c}{650 images \\      80\%  for training, 20\%  for validation, 65 images for test}                                                                  & R2U-Net                                                                                                                                                                                & ACC = 92.15\%                                                                                                                                                                                     \\ \hline
segmentation & TMA cores of PCA tissues                 & 2020 & \cite{Yang-2020-ESA} 133                                      & Q. Q. Yang, et al.                  & 70 TMA cores                                                                                                                                                                             & baseline U-Net                                                                                                                                                                         & F1-score = 83.9\%                                                                                                                                                                                 \\ \hline
segmentation & optic disc                               & 2019 & \cite{Gu-2019-CCE}                                            & Z. W. Gu, et al.                    & \tabincell{c}{650 images\\        50\% for training \\ 50\% for test}                                                                         & \tabincell{c}{CE-Net: Inception-ResNet-V2   block is applied;\\       Atrous convolution is applied;\\      Dense atrous convolution (DAC) blocks is applied} & Overlapping error = 5.8\%                                                                                                                                                                         \\ \hline
segmentation & blood vessels                            & 2020 & \cite{Zhang-2020-DUF}                                         & Z. Zhang, et al.                    & \tabincell{c}{136  RGB images \\      85\%  for training \\ 15\%  for test}                                              & \tabincell{c}{DIU-Net: Inception-res   block  is applied;\\      Dense-Inception block is applied}                                                            & \tabincell{c}{DICE = 95.82\%, \\      JS = 93.38\%, \\      ACC = 96.57\%, \\      SE = 79.67\%, \\      SP = 98.63\%, \\      F1-score = 80.03\%,\\      AUC = 98.02\%} \\ \hline
segmentation & cervical cancer cells                    & 2020 & \cite{Huang-2020-MFM}                                         & Y. F. Huang, et al.                 & 30 images                                                                                                                                                                                & \tabincell{c}{MIRD-Net:  Inception blocks and Residual blocks is   applied;\\      Mini-Inception-Residual-Dense block is applied}                            & DICE = 95.4\%                                                                                                                                                                                     \\ \hline
segmentation & retinal vessel branches and bifurcations & 2020 & \cite{Trimeche-2020-FAC}    140                               & I. Trimeche, et al.                 &\tabincell{c}{30 for  training \\five for validation\\  30 for test } & \tabincell{c}{Residual Inception blocks  is applied;\\      Firesqueeze blocks  is applied}                                                                   & \tabincell{c}{PR = 97\%, \\       RE = 0.96\%,\\       DSC = 96\%,\\      F1-score of = 96\%}                                 \\ \hline

\end{tabular}}
\end{table}

\section{Analysis of Methodology }
\subsection{ Analysis of Several Typical Improvement Methods for U-Net}
According to the survey of improved U-Net: Residual, attention and Inception are used more 
frequently in the improvement of U-Net. Because different datasets are used, each method 
cannot be compared longitudinally, so this paper analyzes from the perspective of the neural 
network itself.

The essence of Residual is a short connection, which means that it adds a skip connection that 
passes by the nonlinear transformation. A pre-made Reset-34 in the feature encoder module is 
used to replace the convolutional layer and a maximum pooling layer. Some previous features can 
be reused, it solves the overfitting problem. Furthermore, because it adds a shortcut mechanism, 
it avoids the disappearance of the gradient and accelerates the network convergence. The papers 
involved in this article 
are~\cite{Patel-2019-CSO,Gomez-2019-DLB,Quan-2016-FAF,Mehta-2018-YJS,Ke-2019-AMU}.

Inception layers are a method of decomposing several convolutions from 1 convolution. The 
Inception layer connects convolutions of different filter sizes ($1 \times 1$, $3 \times 3$, 
$5 \times5$) and the mixing pool layer. Because of the difficulty in choosing the type of 
convolutional layer, Inception is proposed to automate the selection of various layers. The 
papers involved in this article are~\cite{Gu-2019-CCE,Zhang-2020-DUF,Huang-2020-MFM,Trimeche-2020-FAC}.
In deep learning, Inception and Residual are two representative architectures. Inception uses 
different acceptance fields to expand the architecture. On the contrary, Residual uses a 
shortcut connection mechanism to avoid explosions and disappear gradients. Because of its 
existence, the neural network can have thousands of layers. In~\cite{Szegedy-2017-IIA}, the 
Inception-Reset block combines the advantages of Inception and Residual to form a new module.

Attention is also an important way to improve U-Net, a bounding box of the potential region is 
regressed, an attention mask is generated and the attention mask is used as a weighted function. 
It merges discriminative feature maps into the model. Attention gates make the model pay more 
attention to vascular regions. The attention gate further improves the ACC on detailed vessels 
by additionally concatenating attention weights to features before output. The papers involved 
in this article are~\cite{Lian-2018-AGU,Lv-2020-AGU,Mou-2019-CCA,Li-1903-CSA,Jiang-2020-multi,Zhang-2020-PCS,Zhu-2020-SWR,Xiancheng-2018-RBV,Leng-2018-CUF}.

\subsection{ Analysis of the Outstanding Methods in Each Review Task}
In the same dataset, some excellent improved U-Nets are proposed. For example, in the STARE 
dataset task, the best result is obtained by~\cite{Mou-2019-CCA}, where a CS-Net based on the 
self-attention mechanism is designed, so local features and their global dependencies can be 
effectively integrated. In addition, the spatial attention module is introduced, the context 
information is encoded into local features to increase representative capability.

For the Drishti-GS1 dataset segmentation task, the best result is obtained by multi-path recurrent 
U-Net with attention gate~\cite{Jiang-2020-multi}, which introduces the attention gating unit to 
obtain global information of the current input. In addition, maximum pooling and average pooling 
merge operations are introduced to the left and right branches respectively to reduce the error.

In the DRIVE dataset segmentation task, AA-UNet obtains the best segmentation 
results~\cite{Lv-2020-AGU} which generates an attention mask and multiplies it with the feature 
map in the model. The main advantage of this model is that it can pay more attention to the 
blood vessel area. Furthermore, ordinary convolution is replaced by atrous convolution, which 
increases the receiving range and reduces the amount of calculation.

\subsection{The Potential of the Mentioned Methods in this Survey in Other Research Fields}

In addition, this review discussed the U-Net method not only can be applied to microscopic 
images, but also in other image analysis fields. Because of artifacts such as noise, weak edges 
and uneven intensity in Magnetic Resonance Imaging 
(MRI) images~\cite{Piantadosi-2018-BSI,Wang-2019-3UB,Lee-2020-ASO,Rundo-2019-UIS,Kermi-2018-DCN,Chen-2018-AMU}, 
many classic segmentation methods perform poorly. Deep learning neural networks based on U-Net 
can solve this problem.

However, there is no effective way to segment Computed Tomography (CT) images of gallstones. 
In~\cite{Song-2019-UAN}, researchers proposed a new deep learning segmentation model based 
on U-Net with attentional upsampling blocks and spatial pyramid pools, which can give the 
importance of different shapes of target structures in the image. Therefore it is very 
suitable for segmentation of CT 
images~\cite{Roth-2018-AAO,Roth-2018-DLA,Chen-2019-CAS,Saeedizadeh-2021-CTS,Weng-2019-NNA}.

Due to the existence of various ultrasound artifacts, accurate tumor segmentation is still 
a challenge. Therefore, a U-Net segmentation framework based on a deep learning structure is 
often used to segment ultrasound images. A following lists ultrasound image 
analysis~\cite{Almajalid-2018-DOA,Wang-2018-AMS,Li-2019-ASO,Yang-2019-RSO,Amiri-2020-TUI,Amiri-2019-FTU}.

No matter from the selection of datasets, image collection, preprocessing and from the design 
and proposal of an improved U-Net model, the deep artificial neural network summarized in this 
survey paper can provide support for research in other fields as well.

\section{Conclusion and future work}
In this survey paper, U-Net-based microscopic image analysis methods are comprehensively summarized. 
In addition, when summarizing the deep neural network method, the related work is grouped according 
to the type of improvement. In each type of improvement, performances with similar improvements 
idea has categorized similarly.

From classical review works in Sec. 2, U-Net without modification is used in cytology, 
microorganisms, nano-particles, histopathology. Deep learning techniques, especially deep 
convolutional neural networks, have achieved excellent results in the segmentation of microscopic 
images. Pathological images can help patients detect and treat diseases early. A sampling of 
contaminated water sources can separate the microorganisms in it, which is beneficial to protect 
the environment. According to this review Sec. 2, the original U-Net is a very common method, 
but from the review works in Sec. 3 and Sec. 4, the deep learning method based on the improved 
U-Net is the most commonly used. This review works in Sec. 3 and Sec. 4, improved and novel 
network frameworks of U-Net tend to perform well under different datasets.

In the future, there is still room for improvement. First, researchers can combine various new 
microscopic images to develop new U-Net-based models to make them more suitable for new 
microscopic images, such as COVID-19. Secondly, there is no large-scale, full-label, open 
microscopic image dataset with many types and a very large dataset should be established to have 
a positive significance for the exploration of the field of microscopic image segmentation. 
Thirdly, for different medical image microscopy datasets, we should find an improved U-Net that 
is more suitable for them. Since a hospital processes a large amount of data every day, more 
accurate and faster output is also a goal with more potential and value in the future. Finally, 
only a few researchers study the improved U-Net composed of multiple U-Nets, which can better 
realize the flow of information. In the future, this improved U-Net composed of multiple U-Nets 
will be a more potential and valuable research direction, which is a key direction of our future 
work.

\section*{Acknowledgements}
This work is supported by the ``National Natural Science Foundation of China'' (No. 61806047). 
We also thank Miss. Zixian Li and Mr. Guoxian Li for their important discussion in this work.

\section*{Declaration of Competing Interest}
The authors declare that they have no conflict of interest in this paper.

\bibliographystyle{unsrt}       
\bibliography{Wu}   
\end{document}